\documentclass[12pt,preprint]{aastex}
\usepackage{natbib} 
\usepackage{graphicx} 
\usepackage{epstopdf}
\usepackage{amssymb,amsmath}

\newcommand{\HII}{\hbox{H\,{\sc ii}}}
\newcommand{\HI}{\hbox{H\,{\sc i}}}

\newcommand{\NII}{\hbox{[N\,{\sc ii}]}}

\begin{document}

\title{The local radio-IR relation  in M51\\}


\author{G. Dumas and  E. Schinnerer}
\affil{Max-Planck-Institut f\"ur Astronomie, K\"onigstuhl 17, D-69117 Heidelberg, Germany}

\author{F.S. Tabatabaei and R. Beck}
\affil{Max-Planck-Institut f\"ur Radioastronomie, Auf dem H\"ugel 69, D-53121 Bonn, Germany}

\author{T. Velusamy}
\affil{Jet Propulsion Laboratory, California Institute of Technology, Pasadena, CA 91109, USA}

\author{E. Murphy}
\affil{Spitzer Science Center, California Institute of Technology, MC 314-6, Pasadena, CA 91125, USA}


\begin{abstract}
We observed  M51 at three frequencies, 1.4\,GHz (20\,cm), 4.9\,GHz (6\,cm) and 8.4\,GHz (3.6\,cm), with the
VLA and the Effelsberg 100\,m telescope to obtain the highest quality radio continuum images of a nearby spiral galaxy. 
These radio data were combined with deconvolved Spitzer IRAC 8\,$\mu$m and MIPS 24\,$\mu$m images to search for and investigate local changes in the radio-IR correlation. 
 Utilizing wavelet decomposition, we compare the distribution of the
radio and IR emission on spatial scales between 200\,pc and 30\,kpc.  We show that  the radio-IR correlation is not uniform across the galactic disk. It presents a complex behavior with local extrema corresponding to various galactic structures, such as complexes of \HII\ regions, spiral arms and interarm filaments, indicating that the contribution of the thermal and non-thermal radio emission  is a strong function of environment. In particular, the relation of the 24\,$\mu$m and 20\,cm emission presents a linear relation within the spiral arms and globally over
the galaxy, while it deviates from linearity in the interarm and outer regions as well in the inner region, with two different behaviours: it is sublinear in the interarm and outer region and over-linear in the central 3.5\,kpc. Our analysis suggests that the changes in the radio/IR correlation reflect variations of ISM properties between spiral arms and interarm region. The good correlation in the spiral arms implies that 24\,$\mu$m and 20\,cm are tracing recent star formation, while a change in the dust opacity,  'Cirrus' contribution to the IR emission and/or the relation between the magnetic field strength and the gas density can explain the different relations found in the interarm, outer and inner regions.
\end{abstract}


\section{Introduction}

The observed empirical correlation between the total radio and far infrared (FIR) emission of galaxies is very tight \citep[e.g.][]{helou_85, dejong_85}. This relation  holds over  five orders of magnitude in luminosity, from nearby galaxies  \citep[normal spirals, starburst galaxies, including dwarf, irregular and Seyfert galaxies;][]{condon_92, yun_01, bell_03} to distant objects at z$>$3-4 \citep{appleton_04, seymour_09, mark_10a, mark_10b}. 
 This correlation is interpreted as the result of recent massive star formation. Ionized UV photons emerging from young massive stars  ($>$8\,M$_{\sun}$)  heat the surrounding dust, which then reradiates at IR wavelength. These stars evolve to supernovae (SN), whose remnants accelerate cosmic ray (CR) electrons which then emit non-thermal (NT) radio synchrotron emission while propagating through the interstellar magnetic field and losing energy. The radio continuum emission in normal star-forming galaxies at 20   cm wavelength  consists of the dominant NT emission, attributed to synchrotron emission from cosmic ray electrons, and the less important thermal free-free emission emerging directly from  \HII\ regions \citep{condon_92}. At 3.6\,cm, the ratio of thermal fraction to the radio continuum emission can be up to 50\% \citep{fatemeh_07b}. 

The IR radiation is primarily emitted from dust heated by \HII\ regions and  cool dust excited mainly by the interstellar radiation field.  While the thermal radio continuum and warm dust  emission can be both directly associated with \HII\ regions, the IR and NT radio emission mechanisms involve very different physical processes and time scales. Thus the origin and physics of the NT radio-IR correlation remain still under debate. Even though this correlation is still poorly understood, it is used to calibrate the radio emission as a  tracer of recent star formation in local and high redshift galaxies \citep{condon_92, yun_2002, bell_03}. Indeed, the radio continuum is not biased by dust extinction and hence thought to be a good estimator for SF in galaxies, as radio interferometers offer much higher spatial resolution than space-based IR telescopes. The local radio/FIR correlation within galaxies has been investigated \citep{beck_88, hoernes_98,hughes_06,murphy_06,fatemeh_07} in order to probe in detail the origin of this correlation and the mechanisms at work. In particular the determination of the spatial scales on which the radio-IR correlation breaks down can provide strong constraints on the different models explaining this correlation \citep{volk_89,helou_93,niklas_97}. These studies showed that while the peaks of the radio and IR emissions are spatially close \citep{hoernes_98}, the radio map seems to be a smeared version of the IR map \citep{murphy_06},  reflecting the CR electrons' diffusion in the interstellar medium \citep{bicay_90}.

In this paper we investigate the radio continuum emission and its correlation with the IR emission within M51 at high spatial resolution, down to $\sim$200\,pc. Throughout this paper, M51 refers to the interacting galaxy pair while M51a and M51b refer, respectively, to  NGC5194 and the companion NGC5195. The aim of this study is to probe on what spatial scales the radio-IR correlation holds and the role played by the environment for this relation (e.g. arm/interarm regions). In particular, an important point to constrain is how well the radio emission is correlated with SF regions, keeping in mind the ultimate goal of constraining the NT radio continuum as tracer of recent star formation events. M51a is an ideal target to conduct such a study. It is nearby \citep[8.2\,Mpc, which corresponds to $\sim$40\,pc per arcsec,][with $H_0=70$\,km\,s$^{-1}$\,Mpc$^{-1}$]{tully_88}, has well-defined spiral arms and is almost face-on (disk inclination of $\sim$20\,$\degr$). The high disk surface brightness allows one to study in detail arm/interarm variations. It  is also one of the best studied galaxies at almost every wavelengths.

We obtained VLA data at three wavelengths: 20\,cm (1.4\,GHz), 6\,cm (4.9\,GHz) and 3.6\,cm (8.4\,GHz) at high resolution ($\sim$2\arcsec) and high sensitivity (rms $\,<\,25\,\mu$Jy/beam) to probe the distribution of the radio continuum emission across the disk of the galaxy. We compare the radio continuum maps with  SPITZER  MIPS 24\,$\mu$m and IRAC 8\,$\mu$m images and examine the radio-IR correlation down to 200\,pc, using the wavelet transformation and  cross-correlation method  introduced  by \cite{frick_01}. This analysis technique has been successively applied to NGC6946 by \cite{frick_01}, M33 by \cite{fatemeh_07} and the Large Magellanic Cloud by \cite{hughes_06} and allows us to investigate the radio and IR emission and their correlation  as a function  of the different galactic structures.

The paper is organized as follows: In Sect.~\ref{sec:data} we present our radio observation and their data reduction plus the ancillary data used here. We describe the radio continuum emission of M51 in Sect.~\ref{sec:radio_cont}. In Sect.~\ref{sec:analysis} we conduct the structural analysis of the radio and IR images of M51a using continuous 2D wavelet transformation and we investigate the results of this analysis in the context of the radio-IR correlation in Sect.~\ref{sec:radio_ir}. 
\section{Observations and Data reduction}
\label{sec:data}
In the following we describe the datasets used in this paper. An overview is provided in Table~\ref{data}.
\subsection{Radio observations}

 M51 was observed with the Very Large Array (VLA) at 20\,cm (1.4\,GHz) in the A, B, C and D configuration, at  6\,cm (4.9\,GHz) in the B, C and D configuration and at 3.6\,cm (8.4\,GHz) in the C and D configuration. At 3.6\,cm, the primary beam FWHM is 5\farcm 4, thus two pointings were necessary to cover the full disk of M51 with a size of 10\arcmin. The VLA data were taken between 1988 and 2005 as part of several projects, corresponding to a total integration time of $\sim$170\,hrs on-source spread over all three frequencies. Table~\ref{vladata_obs} provides a summary of the different observation runs. For the two most compact configurations (C and D) the  continuum correlator mode was used,  while observations in A and B configuration were done in multichannel continuum mode, using the four-IF spectral line mode. Both setups use two intermediate frequencies (IF). The central frequencies and corresponding bandwidths are listed in Table~\ref{vladata_obs}.
During all observations, the quasar 3C286 was observed for flux calibration. For the 20\,cm observations in C configuration, the quasars 3C48, 3C138 and 3C138 were also available as additional flux calibrators. The different projects used the following sources as phase calibrators: 1335+457, 1349+536, 1409+524, 1219+484 and 137+434. Depending on the frequency and configuration, the phase calibrator was observed every 10\,min to 45\,min.

Data reduction was done using standard AIPS routines. Each day of observation has been calibrated separately. We performed standard flux calibration and atmospheric correction for the uv-data of each observation day. Before and after those calibrations, the data were inspected for RFI (for each IF and polarization) and flagged in consequence. For the data observed in the multichannel correlator mode (20\,cm in A and B configuration and 6\,cm in B configuration), calibration and flagging were done on the pseudo-continuum channel and then applied to the 'Line' data. Bandpass calibration was also performed  on the 'Line' data using the flux calibrator and the pseudo-continuum channel. Finally we extract the source from the calibrated data with the task SPLIT which produces single-source files and applies the calibration. All days of observation in one configuration were then combined together with the task DBCON. At this point, we have one fully calibrated single-source data set for each configuration and each frequency.
 
For each frequency, we combined the fully calibrated uv data in the different configurations. First, we converted the coordinates of the oldest uv data from the B1950 to J2000 coordinates system with the tasks REGRD and UVFIX (20   cm data in C and D configuration and 6\,cm data in D configuration).  Then, the intermediate frequencies being different for the different projects, the IFs  were splitted with the task SPLIT for each configuration, and then recombined with the task DBCON into a single-channel data set. In the case of  uv data obtained in the multichannel continuum
correlator mode  (configurations A and B at 20\,cm and 6\,cm), this step was done after all channels were splitted and recombined
into a single data set. 
  Therefore, at each frequency, a single data set was obtained for all configurations which were then  combined in chronological order (the older observations first) with DBCON to obtain a final uv data set at a given reference frequency. 

 In a next step, the radio maps are produced using the task IMAGR.  As the 20\,cm and 6\,cm data cover a large field-of-view, we first divided the field into 13 and 55 facets, of 4096$\times$4096 pixels and 2048$\times$2048 pixels at 6\,cm and 20\,cm, respectively. Each facet at 20\,cm and 6\,cm (respectively, each pointing at 3.6\,cm) was then imaged, with the multi-resolution CLEAN in AIPS \citep{rich_08, greisen_09}, using several circular Gaussian model sources, in order to optimally deconvolve and image the extended faint emission. At 6\,cm and 3.6\,cm, we choose a number of circular Gaussian models equal to the number of observed configurations: 2 Gaussian models at 3.6\,cm, of width equal to 0\arcsec\  (point source) and 6\farcs 3 (with a flux density cut-off of 35 and 40\,mJy/beam, respectively), 3 Gaussian models  at 6\,cm, of width 0\arcsec , 6\farcs 3 and 20\arcsec\ with flux density cut-offs of 0.05, 0.06 and 0.2\,mJy/beam, respectively. At 20\,cm, 4 Gaussian models of width 0\arcsec , 4\farcs 7, 15\arcsec\ and 47\arcsec\  were chosen (with a respective flux density cut-off of 0.03, 0.3, 3 and 30\,mJy/beam). The widths of the Gaussians are multiple of the robust 0 beam size (respectively 2\farcs4, 2\farcs0 and 1\farcs5 at 3.6cm, 6cm and 20cm respectively), separated by a factor of $\sqrt{10}$, thus giving a factor of 10 between the beam area of successive Gaussians. For the data at 20\,cm and 6\,cm we used respectively a robust weighting of ROBUST$\,=\,1$ and ROBUST$\,=\,0$, and at 3.6\,cm we deconvolved with uniform weighting for maximal resolution (ROBUST$=-4$). After the deconvolution, we  combined the facets/pointings using the task FLATN to construct the final images at 20\,cm, 6\,cm and 3.6\,cm. This task performs also a correction for the primary beam attenuation while combining the two pointings at 3.6\,cm.  During this correction, pixels outside the 10\% power radius of the primary beam were blanked.

At 6\,cm and 3.6\,cm, the largest scale structures detectable (5\arcmin\ and 3\arcmin, respectively) are smaller than the size of the galaxy system (10\arcmin). This implies that at these wavelengths the VLA observations can miss extended emission from M51. To correct for  missing short-spacings, we use the same method as \cite{fatemeh_07b}, which consists of combining the final VLA  CLEANed images with single dish data obtained with the Effelsberg 100m telescope  \citep{fletcher_10}. First, the VLA maps were convolved to a circular beam, 2\arcsec\ at 6\,cm and 2.4\arcsec\ at 3.6\,cm, the corresponding Effelsberg beam size being 180\arcsec\ and 90\arcsec, respectively. We subtracted  strong point sources in the VLA images  to avoid bias in the combined total flux and the VLA and Effelsberg maps were combined in the uv-plane with the task IMERG in AIPS.  The only free parameter of this task is the choice of the uv range of the interferometric data that should be filled by the single-dish values.  We choose the uv range such  that the integrated flux densities of the resulting merged map and of the Effelsberg map are equal. The overlap uv-range is 0.51k$\lambda$ to  0.71k$\lambda$ at 6\,cm  and  
1.01k$\lambda$ to 1.33k$\lambda$  at 3.6\,cm. The VLA point sources were then added back to the short-spacing corrected (SSC) maps. The total flux densities of the resulted SSC and the corresponding single dish maps are in agreement withing 0.1\%. 
Short-spacing correction  is not necessary at 20\,cm, since observation from the compact D configuration of the VLA allows one to detect structures with sizes up of 15\arcmin, larger than M51. 

Finally, we correct the 20\,cm and SSC 6\,cm maps for primary beam attenuation with the task PBCOR, with the same cut-off value of 10\% of the primary beam radius used for the 3.6\,cm image. 
The final 3.6\,cm SSC, 6\,cm SSC and 20\,cm maps corrected from primary beam attenuation have 4096$\times$4096 pixels with a pixel size of 0\farcs 4,  0\farcs 2 and 0\farcs 3 at 3.6\,cm, 6\,cm and 20\,cm, respectively,  and a final resolution of 2\farcs 4$\times$2\farcs 4 at 3.6\,cm,  2\arcsec$\times$2\arcsec\ at 6\,cm and 1\farcs 4$\times$1\farcs 3 at 20\,cm.  The r.m.s. noise level  is similar for the  20\,cm ($\sigma_{noise}$=11\,$\mu$Jy/beam) and 6\,cm data ($\sigma_{noise}$=16\,$\mu$Jy/beam), while it is higher in the case of the 3.6\,cm map ($\sigma_{noise}$=25\,$\mu$Jy/beam).  

\subsection{Spitzer infrared  data}
\begin{figure}
\begin{center}
\includegraphics[width=\textwidth]{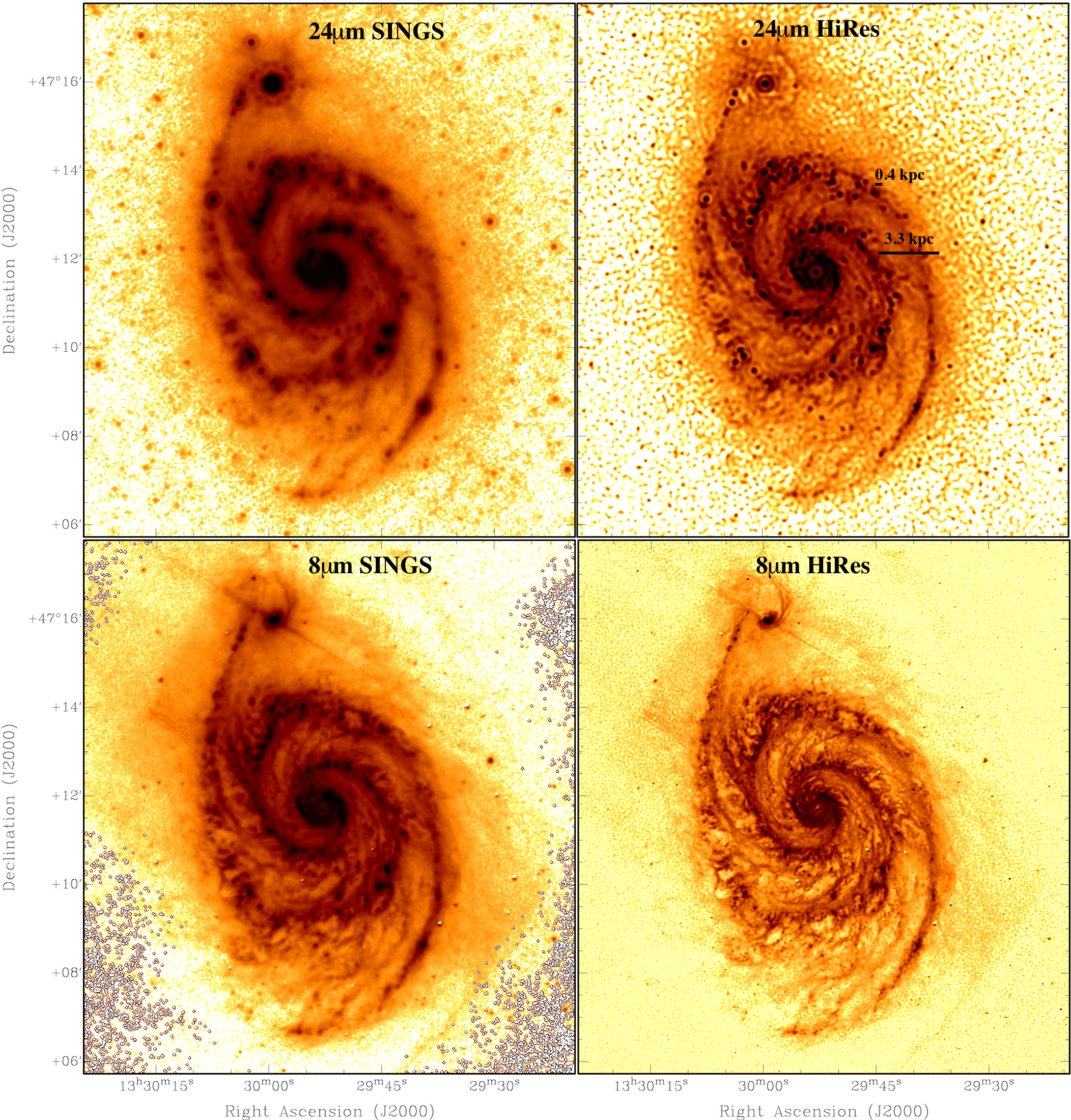} 
\caption{SINGS (left) and HiRes deconvolved (right) images of Spitzer MIPS 24\,$\mu$m (top) and PAH IRAC 8\,$\mu$m (bottom) observations of M51. The horizontal lines in the top right panel correspond to 10\arcsec$\sim$0.4kpc, size of complexes of star forming regions, and to 83\arcsec$\sim$3.3kpc corresponding to the scales encompassing the width of the spiral arms and interarm regions.}
\label{fig:HIRES}
\end{center}
\end{figure}

For comparison with our high resolution radio maps, we used the Spitzer MIPS 24\,$\mu$m map from the fifth Spitzer Infrared Nearby Galaxies Survey (SINGS) data delivery\footnote{http://irsa.ipac.caltech.edu/data/SPITZER/SINGS/} and the 8\,$\mu$m PAH emission map, which corresponds to a stellar continuum subtracted 8\,$\mu$m image, created by \cite{regan_pah06}. 
The description of the observations and reduction can be found in \cite{kennicutt_03} and \cite{sings_2005}. Briefly,  the data reduction of these images used the standard MIPS pipeline \citep{gordon_05}. In addition, some extra steps were performed to improve the image quality. For the MIPS 24\,$\mu$m image, flat-fielding was first performed, hot and dead pixels were removed and latent images from bright sources and the background were subtracted. Cosmic rays were  removed while the mosaic was created.    For the IRAC images additional steps include correction from geometric distortion and rotation, image offsets and bias drift, cosmic ray removal and background subtraction. Finally the stellar continuum was subtracted \citep{regan_pah06}. The final resolution at 24\,$\mu$m is 5\farcs 7 (Fig.~\ref{fig:HIRES}, top left) and it is  2\arcsec\  for the 8\,$\mu$m PAH emission map (Fig.~\ref{fig:HIRES}, bottom left).

As the MIPS point spread function (PSF) has strong secondary lobes and Airy rings which would affect a direct comparison to the radio images, we applied the HiRes deconvolution algorithm \citep{backus_05} to the IR maps. In addition, this deconvolution allows for an increase of the resolution of up to a factor of two in FWHM \citep{velu_08}. Therefore we can analyse our radio and IR data and investigate the correlation between the two without degrading our radio data by convolution with the MIPS PSF as it has been done in a previous study \citep{murphy_06}. HiRes is based on the maximum correlation method (MCM), an extension of the Lucy-Richardson deconvolution algorithm and has been well tested on IRAC as well as MIPS data \citep{backus_05, velu_08}. 
The deconvolution has been  carried out with 50 iterations on the fully reduced SINGS MIPS mosaic at 24\,$\mu$m and on the PAH IRAC 8\,$\mu$m maps, using the point response functions (PRF) provided by the Spitzer Science Center (SSC). The resulting HiRes deconvolved images are shown in Fig.~\ref{fig:HIRES}, along with the initial SINGS images at 24\,$\mu$m  (top left and top right, respectively) and  8\,$\mu$m (bottom left and bottom right, respectively). The HiRes deconvolution provides a FWHM  in the range of 0\farcs 9 - 1\farcs 5 and 2\farcs 2 -2\farcs 6 for the 8\,$\mu$m and 24\,$\mu$m images, respectively, as obtained by fitting Gaussian models to several  point sources in the vicinity of M51.  The spread in the FWHM is caused by a lack of very bright stars in the FOV. 
 We adopt a final resolution of 1\farcs 1$\times$1\farcs 3 at 8\,$\mu$m and 2\farcs 3$\times$2\farcs 4 at 24\,$\mu$m. 

The benefit of applying the HiRes deconvolution is  improved   visualization of spatial morphology by enhancing resolution  and removing the contaminating sidelobes.   Removing  the diffraction lobes in the Spitzer images is more critical than resolution enhancement for our analysis of the radio/IR correlation. 
 The sharper inner edges of the spiral arms become clearly visible in the HiRes image.  The 24\,$\mu$m HiRes map reveals in detail the filamentary structure in the inter-arm regions and allows us to separate some point sources which are blended in the lower resolution SINGS maps. More importantly, the sidelobes of the MIPS PSF have been removed around point sources. This is evident in the case of the companion and for strong point sources in the disk of M51a.  

HiRes deconvolution is most effective in removing diffraction spikes  in the case of isolated point sources located in a zero background.  However, when the point sources are embedded in a high background  (like the nucleus of M51a  and bright point sources in the spiral arms), point sources show a residual ringing at a very low level of 0.2\% to 0.5\%,  although the diffraction spikes are removed.  This level is significantly higher than the nominal diffraction residual around isolated bright point sources.  For comparison the ring around the nucleus of M51b is weak at the 0.2\% level.   Therefore the nucleus of M51a and the region around the nucleus of M51b will be masked in the  correlation analysis applied in the following sections.
 We verified that the flux has been conserved during the deconvolution by comparing the total flux of the initial and deconvolved maps, as well as the flux within individual regions. For this second test, we used  13\arcsec\  diameter apertures centered on bright \HII\ regions as defined by \cite{kennicutt_2007}. The total flux and the fluxes within the apertures centered on the bright \HII\ regions are conserved, with a difference of about 0.7\% between the maps and the rms noise level of the final  HiRes maps are about 0.0067\,MJy/sr at 24\,$\mu$m and 0.0069\,MJy/sr at 8\,$\mu$m.

\subsection{H$\alpha$ data}

The H$\alpha$ image used in our study is the H$\alpha$ emission-line map with the underlying stellar continuum subtracted as presented by \cite{sings_2005}.  An average constant correction factor has also been applied across the image to correct for the contribution of the two \NII\ emission lines \citep{sings_2005}. 
It was obtained on 2001 March 28 at the 2.1\,m KPNO telescope as part of the SINGS Legacy Survey. The measured FWHM of the PSF is 1\farcs 9 comparable to the resolution of our radio continuum data. In addition to standard reduction procedures, correction for the vignetting effect in the southern region of the FOV was applied to these data, by comparing flux measurement of stars in the vignetted side of the mosaic with the same stars in the clean side. Final flux calibration was carried out using standard star observations. The final map, without any further background subtraction applied, is shown in Fig.~\ref{fig:radio}.  The median value of dust extinction in H$\alpha$ is $A_V\,\sim\,2.8$ in $\HII$ regions \citep{sings_2005, scoville_01} and $A_V\,\sim\,2$ in diffuse gas \citep{scoville_01} and it increases towards the center \citep{sings_2005}. Dust absorbs the UV and optical light of young stars and re-emit in IR. Therefore, dust extinction hampers greatly any SFR estimation from optical and UV wavelengths.

\section{M51's radio continuum as seen by VLA at 100\,pc resolution}
\label{sec:radio_cont}

\subsection{Radio morphology}

\begin{figure}
\begin{center}
\includegraphics[width=\textwidth]{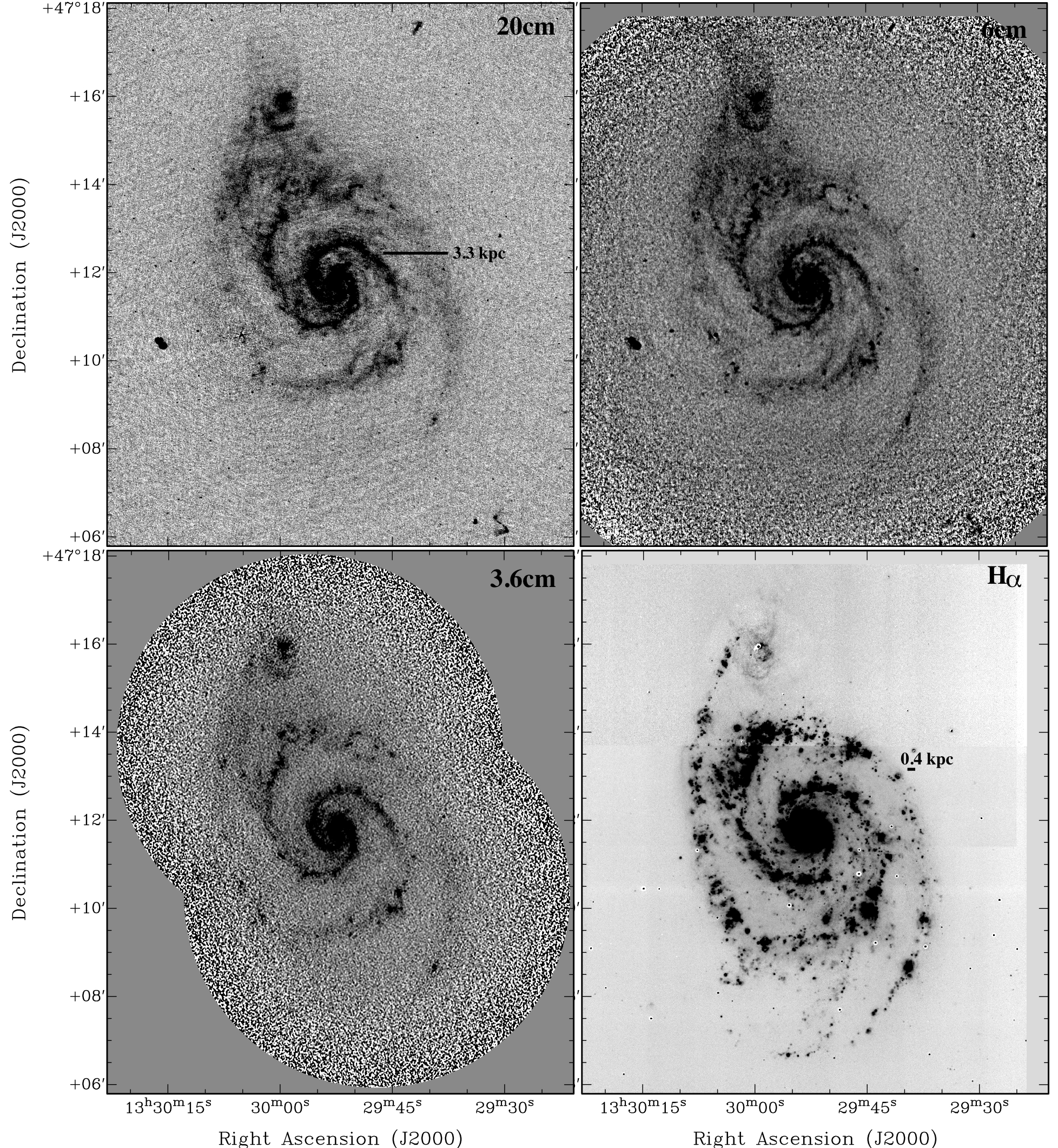} 
\caption{Radio continuum images at 20\,cm (top left), 6\,cm (top right) and 3.6\,cm (bottom left) and the H$\alpha$ map (bottom right). The data at 3.6\,cm and 6\,cm have been short-spacing corrected using single dish data from the Effelsberg 100\,m telescope and all radio maps have been corrected from primary beam attenuation. The resolution of the maps is given in Table~\ref{data}. The horizontal lines  correspond to 10\arcsec$\sim$0.4kpc (bottom right panel), size of complexes of star forming regions, and to 83\arcsec$\sim$3.3kpc corresponding to the scales encompassing the width of the spiral arms and interarm regions (top left panel).}
\label{fig:radio}
\end{center}
\end{figure}

Fig.~\ref{fig:radio} presents our final short-spacing corrected radio continuum maps. These high resolution radio images of M51 show unprecedented details. 
 Many structures are revealed from point sources in the spiral arms to faint extended filaments in the inter-arm and outer regions.  While the general distribution of the 20\,cm continuum, dominated by synchrotron emission, is fairly similar to the morphology seen in the 3.6\,cm and 6\,cm emission (a mixture of synchrotron and thermal radio emission), the contrast of  individual structures is changing from one frequency to another. 

The spiral arms are bright and well defined at all wavelengths, yet faint extended emission is hardly detected at 3.6\,cm. Within the arms, compact bright sources are present at all three frequencies. They mostly correspond to \HII\ regions and SNRs.
The inter-arm regions present  filamentary structures as well as fainter extended smooth emission at 20\,cm and 6\,cm. Those structures are not seen at 3.6\,cm, due to the higher noise of the 3.6\,cm data and its intrinsically higher faintness as evidenced by its spectral index \citep[$\alpha\,\sim\,-0.9$, with $S_\nu\,\propto\,\nu^\alpha$,][]{fletcher_10}. 
Distinct 20\,cm radio continuum emission is present between the northern arm of M51a and its companion. This region is less bright in the 6\,cm map and is hardly evident in the 3.6\,cm map. It is consistent with synchrotron emission that is enhanced by the ongoing interaction of NGC5195 with M51a (see Schinnerer et al. in prep. for a detailed analysis of this region). In particular an interesting feature in this region is the  arc shaped structure south of M51b, also observed in  H$\alpha$ emission \citep[][and Fig.~\ref{fig:radio}]{greenawalt_98}. 
North of NGC5195, extended faint emission is present at 20\,cm and 6\,cm. 
 In the 20\,cm and 6\,cm images, one background radio galaxy with strong radio lobes is observed east (RA~$=13^h30^m16.^s$; $\delta=+47\degr10\arcmin22\arcsec$) of M51. This object is brighter at 20\,cm than 6\,cm, which implies a steep spectrum source. It is not seen in the 3.6\,cm map since it lies outside the mosaic.

\begin{figure}
\begin{center}
\includegraphics[width=12cm]{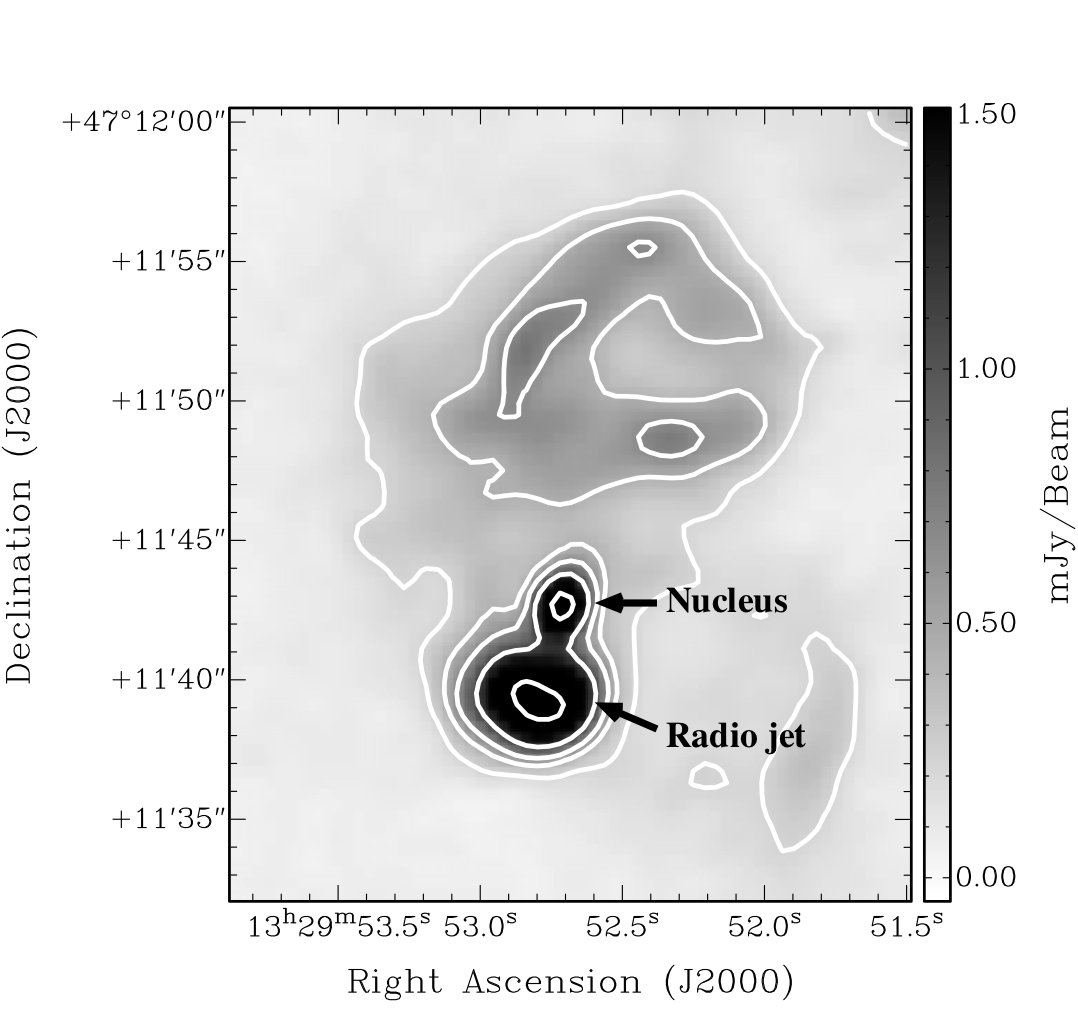} 
\caption{Radio continuum image at 20\,cm of the nuclear region of M51. Contour levels are at 20$\sigma$, 40$\sigma$, 60$\sigma$, 100$\sigma$ and 200$\sigma$, with  $\sigma\,=\,11\,\mu$Jy\,beam$^{-1}$. The positions of the nucleus and radio jet are labeled.}
\label{fig:20centre}
\end{center}
\end{figure}

M51a hosts a type  2 active nucleus. The inner kiloparcsec region observed at 20\,cm is shown in Fig.~\ref{fig:20centre}. The nucleus is unresolved and located at RA~$=13^h29^m52.^s7$; $\delta=+47\degr11\arcmin42\farcs5$ (J2000), in good agreement with the position reported by \cite{crane_92} at 6\,cm. In addition to the radio source, two main nuclear features are observed: a ring-like structure 5\arcsec\ north of the nucleus, thought to be an outflowing bubble, related to the nuclear activity \citep{maddox_07} and  bright compact emission south of the nucleus corresponding  to the edge of the radio jet, identified by \cite{crane_92} at 6\,cm. The jet itself, 3\arcsec\ long, is hardly resolved at our resolution (1.4\arcsec\ at 20\,cm) as seen in Fig.~\ref{fig:20centre}.

\subsection{Radio fluxes and scale lengths}

Using the task IRING we calculated the total integrated flux in the primary beam corrected maps. We used rings centered on the AGN location and assumed a position angle of 170\degr\ and a disk inclination of 20\degr. We found total fluxes of $0.3\pm0.2$\,Jy at 3.6\,cm, $0.6\pm0.2$\,Jy at 6\,cm and $1.4\pm0.1$\,Jy at 20\,cm (including the companion M51b, see Table.~\ref{radio_prop}).  These fluxes are in good agreement with previous single dish and interferometric observation. Indeed at 3.6\,cm \cite{klein_81} 
derive a flux of $0.306\pm0.026$\,Jy, at 6\,cm \cite{israel_83} report a flux of $0.525\pm0.033$\,Jy and at 20\,cm \cite{segalovitz_77} give a flux of $1.5\pm0.1$\,Jy. The global spectral index is $\alpha\,=\,-0.9$ (where the radio flux is expressed as $S_\nu \propto \nu^{\alpha}$), in good agreement with the results of \cite{fletcher_10}.

\begin{figure}
\begin{center}
\includegraphics[height=\textheight]{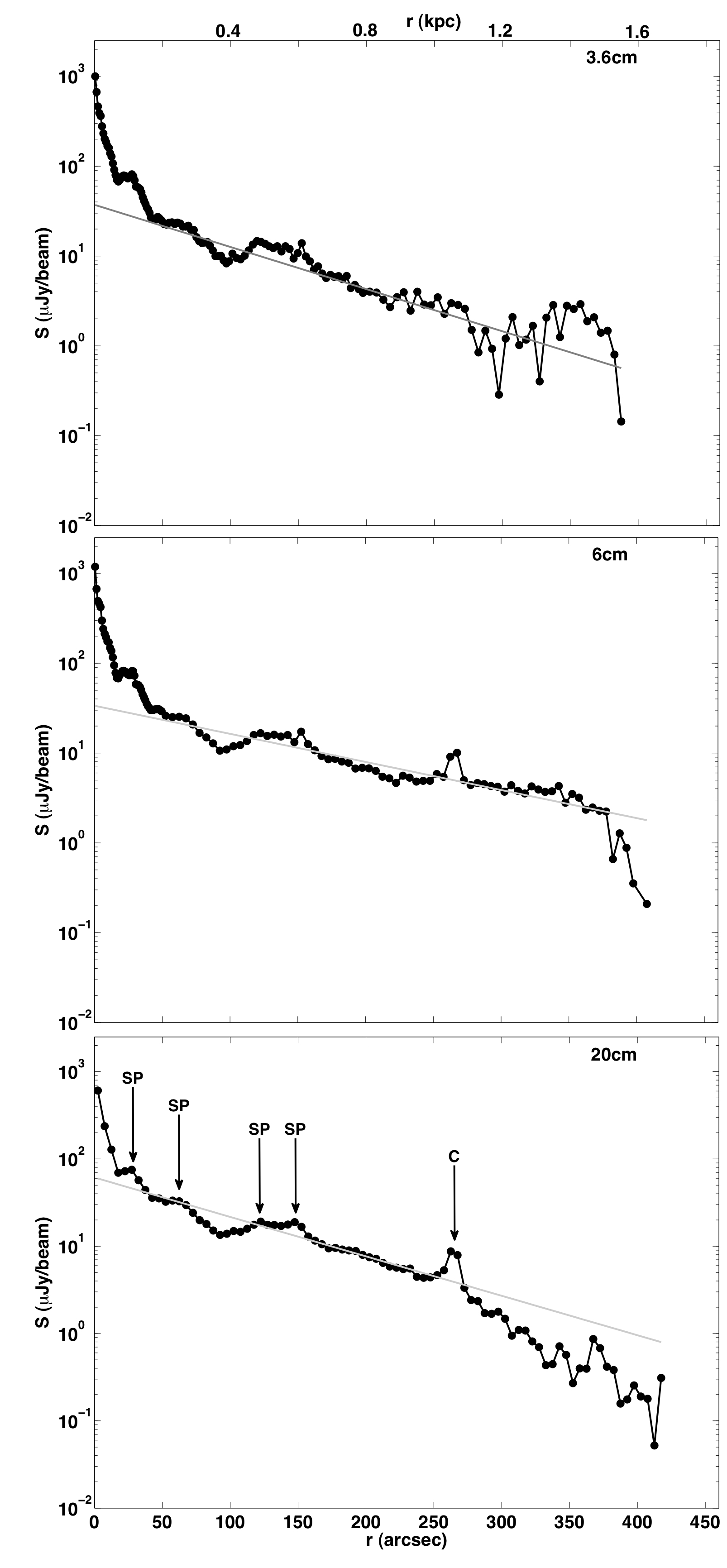} 
\caption{Radial profiles of the radio flux at 3.6\,cm (top), 6\,cm (middle) and 20\,cm (bottom). The straight lines correspond to an exponential fit of the profiles. The peaks at 25\arcsec , 67\arcsec , 120\arcsec\ and 147\arcsec\ correspond to the spiral arms (marked by 'SP'), the companion NGC5195 is seen as a peak at 250\arcsec (marked by 'C').}
\label{fig:radial_profile}
\end{center}
\end{figure}

Within each tilted ring, we also derived the azimuthal average  radio flux. Fig~\ref{fig:radial_profile} presents the radial profiles at 3.6\,cm, 6\,cm and 20\,cm, from 2\farcs 5$\,\approx',$100\,pc to 400\arcsec$\,\approx\,$16\,kpc, encompassing the entire extent of M51.  The azimuthal averaged radial profiles are similar at all three wavelengths. They show several peaks at the same location: 25\arcsec , 67\arcsec , 120\arcsec\ and 147\arcsec\ (1\,kpc, 2.7\,kpc, 4.8\,kpc and 5.9\,kpc). Those radii correspond to the inner and outer spiral arms. One difference between our three maps is that  the profile at 20\,cm and 3.6\,cm decrease for radii R$\,>\,260\arcsec\,\approx\,10.5$\,kpc and R$\,>\,220\arcsec\,\approx\,9$\,kpc, respectively,  while at 6\,cm the intensity remains constant at large radii. The radio profiles at 20\,cm and 6\,cm present  a peak at about 250\arcsec\ (10\,kpc), which is not seen at 3.6\,cm. This peak corresponds to NGC5195. It does not show up in the 3.6\,cm radial profile due to increase of noise at these large radii in the 3.6\,cm map.

We fitted the radial profiles  for radii between 50\arcsec\ and 280\arcsec\ (Fig.~\ref{fig:radial_profile}) with exponentials of the form $I_0\,\times\,\exp{\left(-\frac{r}{h}\right)}$, where $h$ is the exponential radial scale length of the profile. The corresponding scale lenghts are 3.8\,kpc$\,\pm\,$0.2\,kpc at 3.6\,cm, 5.6\,kpc$\,\pm\,$0.4\,kpc at 6\,cm and 4.5\,kpc$\,\pm\,$0.5\,kpc at 20\,cm (see Table.~\ref{radio_prop}).
Our values are lower than the scale length of the molecular gas disk \citep[6.5~kpc;][]{Hera_m51_II} but the scale length at 6\,cm is in good agreement with the dust scale length \citep[5.4\,kpc;][]{meijeirink_05}. At 3.6\,cm the smaller scale length corresponds to the thermal emission, concentrated close to \HII\ regions. We find a smaller scale length at 20\,cm than at 6\,cm, suggesting  that we may miss a very extended smooth component, consisting of old CR electrons, emitting synchrotron emission. Although our VLA data are theoretically sensitive to structures up to 15\arcmin\ (about 35\,kpc), single dish observations seem to be needed to correctly map the largest scale disk at 20\,cm. For comparison, NGC6946 presents a scale length of 4.0\,kpc at 20\,cm (non-thermal emission) and 4.6\,kpc at 6\,cm \citep{walsh_02} while in M33, the scale lengths are much smaller: 2.7\,kpc at 20\,cm, 2.3\,kpc at 6\,cm and 1.3\,kpc at 3.6\,cm \citep{fatemeh_07b}. The missing flux at 20\,cm does not pose a problem for our wavelet analysis, as it only affects spatial scales larger than $\sim$400\arcsec. We verified this by comparing the wavelet analysis of the 6\,cm images with and without short spacing correction.

\section{Structural  Analysis of the Continuum Morphology}
\label{sec:analysis}

We want to investigate and quantify how well the structures in M51a at different spatial scales (spiral arms, filaments in the inter-arm regions, clumps in the spiral arms...) correlate between the different continuum emission maps, in particular the radio and IR maps. In order to study the correlation of structural scales at different wavelengths, the wavelet analysis is a powerful tool \citep[see e.g.][]{frick_01,hughes_2006,fatemeh_07}, as it allows for a comparison of compact (e.g. \HII\ regions), intermediate (e.g. spiral arms) and large structures. The mathematical formalism of the wavelet analysis (definitions, energy conservation and image reconstruction) is described in detail in Appendix~\ref{app:wave}. In this section, we present the wavelet analysis of our multiwavelength data of M51. 

\subsection{Wavelet Analysis of M51}
\label{sec:wave_m51}

To apply the wavelet analysis to our data, we first convolve our radio and IR maps to a common resolution of 2\farcs 4 (which is the resolution of our HiRes 24\,$\mu$m and radio 3.6\,cm images).  The maps were regridded to the same geometry with a spatial sampling of 0\farcs 9 and centered at  RA~$=13^h29^m52.^s7$; Dec~$=+47\degr11\arcmin42\farcs5$ (J2000), the location of the nucleus of M51a. Finally we cut the images to a common field of view of 12\arcmin$\times$12\arcmin\ (799$\times$799 pixels). 
We subtracted five bright radio sources as they dominate the wavelet spectra at small spatial scales and therefore mask the contribution of fainter features. These sources are  the two galactic nuclei (including the jet and bubble related to the AGN of M51a and the surrounding of the nucleus of M51b) and three luminous radio sources at the coordinates RA~$=13^h29^m30.^s5$; Dec~$=+47\degr12\arcmin50\farcs5$, RA~$=13^h29^m51.^s6$; Dec~$=+47\degr12\arcmin08\arcsec$ and RA~$=13^h30^m05^s$; Dec~$=+47\degr10\arcmin35\farcs7$ (J2000). The latter three sources correspond to the sources 1, 37 and 104 in the catalog of \cite{maddox_07}. As they exhibit steep  radio spectral indices and no optical or X-ray counterparts have been detected, they are very likely background radio galaxies \citep{maddox_07}. 

  In order to avoid any boundary conditions while scanning the dynamical range of spatial scales with the wavelet decomposition, the final maps are 3 times larger than the  original maps (2397$\times$2397 pixels). We developed a new tool to calculate the wavelet transform using MatLab and the YAWTb toolbox (Yet Another Wavelet Toolbox\footnote{http://rhea.tele.ucl.ac.be/yawtb/}), to be able to handle such large maps. Within this toolbox, 
the continuous 2D wavelet transform is implemented in terms of Fourier transforms, using the FFT algorithm within MatLab:
\begin{equation}
W(a,\boldsymbol{x}) =  \int_{-\infty}^{\infty}\int_{-\infty}^{\infty} \hat{f}(\boldsymbol{k}) \hat{\psi^{*}}( a\boldsymbol{k})e^{2i\pi \boldsymbol{k}.\boldsymbol{x}} d\boldsymbol{k}
\end{equation}
 where $f$ is the 2D function to be analysed and $\psi$ is the wavelet function. $\hat{f} $ and $\hat{\psi}$ are the Fourier transforms of $f$ and $\psi$, respectively.

The choice of the analysing wavelet depends on the data and the goals of the analysis. Here  we choose the Pet Hat function introduced by \cite{frick_01} which allows a  better separation of scales than most commonly used wavelets (e.g. Morlet, Haar or Mexican Hat wavelets). The Pet Hat function  is defined in the Fourier space as:

\begin{equation}
\hat{\psi}(\boldsymbol{k}) = \left\{ \begin{array}{ll}
\cos^2 \left(\frac{\pi}{2}\log_2 \frac{k}{2\pi}\right), & \pi \leq k \leq 4\pi \\
0, & k> 4\pi,  k<\pi
\end{array} \right.
\end{equation}

where $k=|\boldsymbol{k}|$.

While the wavelet analysis is robust to noise, compared to Fourier analysis \citep{frick_01,fatemeh_07}, it is not obvious how exactly noise affects the smallest scales probed at low signal-to-noise ratio (S/N).  In our radio data, the noise at the map edges is increased due to the primary beam correction of the interferometric observations. Since we have added single dish data to our 6\,cm and 3.6\,cm VLA maps for short spacing correction, blanking pixels of values lower than, for example, 2$\sigma_{rms}$ removes also a lot of this intrinsically faint, but real, emission.  This is not the case for the 20\,cm and the IR and optical maps, but for consistency we applied the wavelet analysis to our data without blanking pixels with low S/N.

 We tested the influence of noise by gradually adding Gaussian noise  to our 20\,cm map (which has the best signal-to-noise ratio of the radio data)  from 11\,$\mu$Jy/beam (noise level of the 20\,cm map) to 25\,$\mu$Jy/beam (noise level of the 3.6\,cm map) in steps of 1\,$\mu$Jy/beam. The description of the noise effect on the wavelet analysis can be found in Appendix~\ref{noise_effect}.  The conclusion of our test is that noise affects the wavelet analysis on small spatial scales: $a\,<\,18\arcsec$ ($a\,<\,10\arcsec$) at the noise level of the 3.6\,cm map (the 6\,cm map, rms$\,\approx\,16\,\mu$Jy/beam).  Therefore, we consider the wavelet spectra of 3.6\,cm and its correlation with the other wavelengths only on scales larger than 18\arcsec, while this limit is set to 10\arcsec\ for the spectrum and cross-correlation of the 6\,cm data.

For the final analysis, we decompose the radio, IR and H$\alpha$ maps 
into 50 spatial scales between 5\arcsec\ (twice the resolution, except for the radio maps at 3.6\,cm and 6\,cm, see above) and 650\arcsec\ ($\approx\,$26\,kpc, maximum extent of M51a) to encompass all structures within the galaxy with enough spatial sampling to compare the morphology at different wavelengths.

\subsection{Dominant spatial scales as revealed by wavelets}
\label{sec:results}

In this section we describe the results of the wavelet analysis of M51a's radio, IR and optical maps.  Table.~\ref{wave_minmax} summarizes the interesting scales found in the wavelet spectra (labeled with a 'S', second and third column) or cross-correlation between two images (labeled with a 'X', fourth and fifth column). By construction, the steps between the spatial scales are about 10\% of $a$: $\delta a \sim \frac{a}{10}$,  increasing toward larger scales. Moreover as it is described in details in the following, there are shifts of the interesting scales between the wavelengths. Therefore, for a better clarity of   Table.~\ref{wave_minmax}, the errors on $a$ listed in the first column are the maximum between the steps and the spatial shift of the local extrema discussed below between the different emissions.   

\subsubsection{Wavelet spectra}
\label{sec:wavespectra}
The wavelet spectrum, $M(a)$, represents the distribution of the emitting power as a function of the spatial scale $a$: $$ M(a) = \int_{-\infty}^{\infty}\int_{-\infty}^{\infty} |W(a,\boldsymbol{x})|^2 d\boldsymbol{x} $$ where $W(a,\boldsymbol{x})$ is the wavelet decomposition map at the spatial scales $a$ \citep[see][and Section~\ref{sec:wave_spec} in this paper for a detailed mathematical definition]{frick_01}. Towards larger scales, a smooth increase in $M(a)$ is expected if most of the emission is coming from diffuse structures, and a decrease if strong point-like or compact structures are present. Peaks in the distribution of  $M(a)$ indicate dominant scales in the morphology.  

The wavelet spectra derived for various continuum and line maps of M51 are presented in Fig.~\ref{wave_spec}. For a better clarity of this figure, the spectra have been scaled by powers of 10, avoiding any confusion. None of the spectra increases smoothly towards larger scales indicating the absence of a dominant diffuse disk. The only general increase is visible at 20\,cm, while the wavelet spectrum at 24\,$\mu$m decreases and the spectra of the other frequencies remain roughly flat.  These general trends are  accompanied by some fluctuations, which indicates that prominent scales, and thus distinct morphological structures, exist.  Local maxima and minima in the wavelet spectra of our different maps are listed in Table.~\ref{wave_minmax}, in the second and third columns respectively.

All three wavelet spectra of the radio data exhibit similar substructures. The 3.6\,cm (6\,cm) spectrum starts at 18\arcsec\ (10\arcsec) as the noise dominates at smaller scales
 (see Appendix~\ref{noise_effect}).  The fluctuations in the three radio spectra can be interpreted as several local extrema, roughly located at the same spatial scales. All three
 radio wavelet spectra show a small bump at $a\,\simeq\,20\arcsec$ (0.8\,kpc, corresponding to complexes of star forming regions and structures within the spiral arms).  The
 second local maximum can be attributed to  the width of  the spiral arms seen in radio and a third maximum appears on a larger scale where the resolved structures include the
 inner disk of M51a, NGC5195 and the outer spiral arms. Interestingly, these two maxima appear at slightly  larger scales at 20\,cm (respectively at $a\,\simeq\,55\arcsec
\sim\,2.2\,$kpc and $150\arcsec\,\sim\,6\,$kpc) than for the other two radio wavelengths (at about $45\arcsec\,\sim\,1.8\,$kpc and $\sim\,135\arcsec\,\sim\,5.5\,$kpc), 
indicating wider spiral arms and a larger disk at 20\,cm likely due to the propagation of  cosmic ray electrons.

      The most dominant scale in our data is found at $a\,=\,10\arcsec$  (0.4\,kpc) where the 24\,$\mu$m emission is considerably strong. No dominant scale is seen in the 8\,$\mu$m (PAH) or the H$\alpha$ emission. The latter could indicate a strong effect of extinction in the H$\alpha$ emission by dust flattening the spectrum on small scales, (since we would expect a small-scale dominance in the H$\alpha$ emission, similar to the one found for the 24\,$\mu$m emission).   Small scales ($a<$20\arcsec) contribute more to the emission at the IR and optical wavelengths than in the radio continuum.  Although the IR emission at 8\,$\mu$m and 24\,$\mu$m is mostly related to warm dust, their spectra are strikingly different. While the 24\,$\mu$m emission is strongest at 10\arcsec, the PAH 8\,$\mu$m spectrum is approximately flat over our spatial scales range.  This reflects the fact that the 24\,$\mu$m emission is more strongly peaked on star forming regions than the PAH 8\,$\mu$m.  The 8\,$\mu$m spectrum reaches a very weak local maximum at about 16\arcsec$\,\sim\,640$\,pc. This weak maximum could trace the PAHs illuminated by young and massive stars. However PAHs emission arising from PDRs surrounding the \HII\ regions \citep{helou_04, murphy_06, bendo_06, gordon_08} is not predominant and the contribution of PAHs heated by the interstellar radiation field on larger scales is significant, as shown in previous studies \citep[e.g.][]{bendo_08}.  Moreover,  dust emission may be important in regions of intense radiation field at 8\,$\mu$m \citep{draine_07} which could also explain the significatn contribution of the large scales in the wavelet spectrum of this wavelength.
           The 8\,$\mu$m spectrum shows a local maximum at $a\,\simeq\,45\arcsec$ ($\sim\,1.8$\,kpc) similar to the radio 3.6\,cm and 6\,cm emission spectra (the wavelet spectrum at 20\,cm reaches a maximum at 55\arcsec\ ). The  H$\alpha$ and 24\,$\mu$m wavelet spectra present no minimum at this scale, and decrease smoothly out to $a\,\simeq\,200\arcsec$ ($\sim\,8\,$kpc). This indicates that while structures corresponding to the spiral arms are clearly visible at 8\,$\mu$m they do not dominate the morphology of the H$\alpha$ and 24\,$\mu$m emission. Both IR and optical wavelet spectra have a weak maximum around scales of 120-130\arcsec\ ($\sim\,4.5\,$kpc) corresponding to structures encompassing M51a inner disk, the outer spiral arms and the companion NGC5195.  Finally a strong minimum at 200-220\arcsec ($\sim\,8.4\,$kpc) is present in both IR and optical wavelet spectra at the same location than the minima in the radio wavelet spectra. This is moreover the only obvious feature in the H$\alpha$ spectrum. This strong local minimum in all wavelet spectra may reflect the good contrast between the disks of the two interacting galaxies on this spatial scales ($\sim\,8.4\,$kpc). This minimum is also weaker in the wavelet spectra of the radio emission, revealing a diffuse radio structure encompassing both M51a and M51b and interestingly the 3.6cm wavelet spectrum reaches this local minimum at larger scales ($\sim\,300\arcsec \sim\,12\,$kpc). 
           The locations of the local extrema in the optical and IR wavelet spectra are similar to those  of the extrema present in the radio spectra, albeit at systematically shorter spatial scales: the sizes of the corresponding structures are 10\% smaller  than those in the radio maps. This shift between the sizes of optical/IR and radio structures is likely reflecting the diffusion of the CR electrons into the ISM which makes the radio structures appear wider and larger than the IR and optical counterparts. 

Finally, at all wavelengths the wavelet spectra reach a maximum  around 600\arcsec\ ($\sim\,24$\,kpc) and remain almost constant at this maximum for larger scales. On these scales, the wavelet probes the extended galactic disk of M51a. This maximum is reached at smaller scales in the IR and H$\alpha$  than in  the radio maps, the difference is about 5\,kpc. This shows a larger extent of the disk in radio than in IR and H$\alpha$, again indicating propagation of CR electrons producing the radio emission. This finding is in good agreement with \cite{murphy_06, murphy_2008} who showed that radio images can be interpreted as a smeared version of the IR maps.

\subsubsection{Wavelet cross-correlations}
\label{sec:crosscorr}
The wavelet spectrum allows us to identify the dominant energy structures in our data and to qualitatively compare them. The wavelet cross-correlation is a useful method to compare different images as a function of spatial scales as shown in several studies \citep[e.g.][]{frick_01,hughes_2006,fatemeh_07}.  Indeed, while standard cross-correlation analysis, such as a pixel-to-pixel correlation, can be dominated by bright extended regions or large-scale structures, the wavelet cross-correlation $r_w$ allows for the analysis of a scale-dependent correlation  between two images.  It is also very sensitive to any  differences between two images even on spatial scales where the wavelet spectra appear similar. Such differences could be, e.g. an azimuthal off-set between  spiral arms or a systematic shift of structures of similar size. The definitions and equations used here are detailed in Sect~\ref{sec:wave_corr}. Following \cite{frick_01} we consider the correlation between two images to be significant for values higher than $\rm {r_w > 0.75}$. Fourth and fifth column of Table.~\ref{wave_minmax} list the local maxima and minima  in the wavelet cross-correlation of our different maps. 

There is significant correlation between the radio images at all scales larger than $\sim\,20\arcsec$ (Fig.~\ref{cross_corr_radio}). At all scales, the best correlation occurs between the 20\,cm and 6\,cm maps while the weakest is between 20\,cm and 3.6\,cm particularly for scales  $a\,<\,140\arcsec$.  Since the observed radio continuum is a combination of thermal (free-free) and non-thermal (synchrotron) emission, the contributions from thermal and non-thermal emission change as a function of wavelength as well as location in the galactic disk, the fraction of non-thermal emission being more and more important at long wavelength as well as on large structures. This can explain the differences between the wavelet correlations shown in Fig.~\ref{cross_corr_radio}.  Two local maxima and two local minima are present in these cross-correlations at similar spatial scales and are especially pronounced in the cross-correlations involving the 3.6\,cm data. The locations of the first local maximum and minimum correspond well to those seen in the wavelet spectra of the radio maps and corresponds to the width of the spiral arms (about 2\,kpc)  and the size of the structures formed by the combination of the spiral arms and the interarm region ($\sim$3.5\,kpc) .  The local minima could be attributed to different slopes and displacements in the extrema of the corresponding wavelet spectra (Fig.~\ref{wave_spec} and Table~\ref{wave_minmax}) again caused by more diffuse morphology of the longer wavelength radio emission (due to a higher contribution from synchrotron emission). 

Now we study the wavelet correlation between the radio continuum at 20\,cm, 6\,cm and 3.6\,cm and the IR/optical emission (Figures~\ref{cross_corr_IR} and \ref{cross_corr_ha}, respectively).  First we notice that all the correlation coefficients decrease globally towards smaller scales and this decrease is particularly steep from $a\,=\,40\arcsec$ downwards, which corresponds to the width of the spiral arms. The correlation of the IR emission at 24\,$\mu$m is higher with the radio continuum at 3.6\,cm than at 6\,cm and 20\,cm on spatial scales smaller than 100\arcsec\ (Fig.~\ref{cross_corr_IR}). This is expected, as the radio continuum at 3.6\,cm contains less synchrotron contribution than at 6\,cm or 20\,cm, the high correlation between 3.6\,cm and IR data reflects the common origin of the thermal radio and IR emission, both being powered by ionizing young stars in the \HII\ regions. Local minima   and maxima  are present in the  three correlations between radio and IR emission, appearing at similar spatial scales as the local extrema in the cross-correlation between the radio wavelengths themselves. Finally, the correlation between 24\,$\mu$m and 8\,$\mu$m PAH emission is the strongest correlation involving IR or optical emissions, at all spatial scales. It reaches a weak local minimum at $a\,=\,245\arcsec$ ($\sim\,7.3$\,kpc) which is the location of the local maximum of the cross-correlation between 24\,$\mu$m and 3.6\,cm.

The correlations including H$\alpha$ emission decrease more rapidly than the corresponding correlations with IR emission (Fig.~\ref{cross_corr_ha}). They all present a local maximum at about 135\arcsec\ ($\sim$5.7\,kpc) and minimum at 200\arcsec\ ($\sim$8.2\,kpc, this local minimum is reached at slightly larger scale for the correlation with 3.6\,cm). The wavelet cross-correlations between H$\alpha$ and the IR emissions at 24\,$\mu$m and 8\,$\mu$m are very similar. As in the case of the 24\,$\mu$m emission (see Fig.~\ref{cross_corr_IR}), the  wavelet cross-correlations between H$\alpha$ and the radio emissions are lower at 20\,cm than at 6\,cm and 3.6\,cm at all scales $a\,<\,200\arcsec$. 

\section{The local  radio-IR correlation within M51}
\label{sec:radio_ir}

\subsection{ Characteristic scales }
\label{scales}
\begin{figure}
\begin{center}
\includegraphics[width=\textwidth]{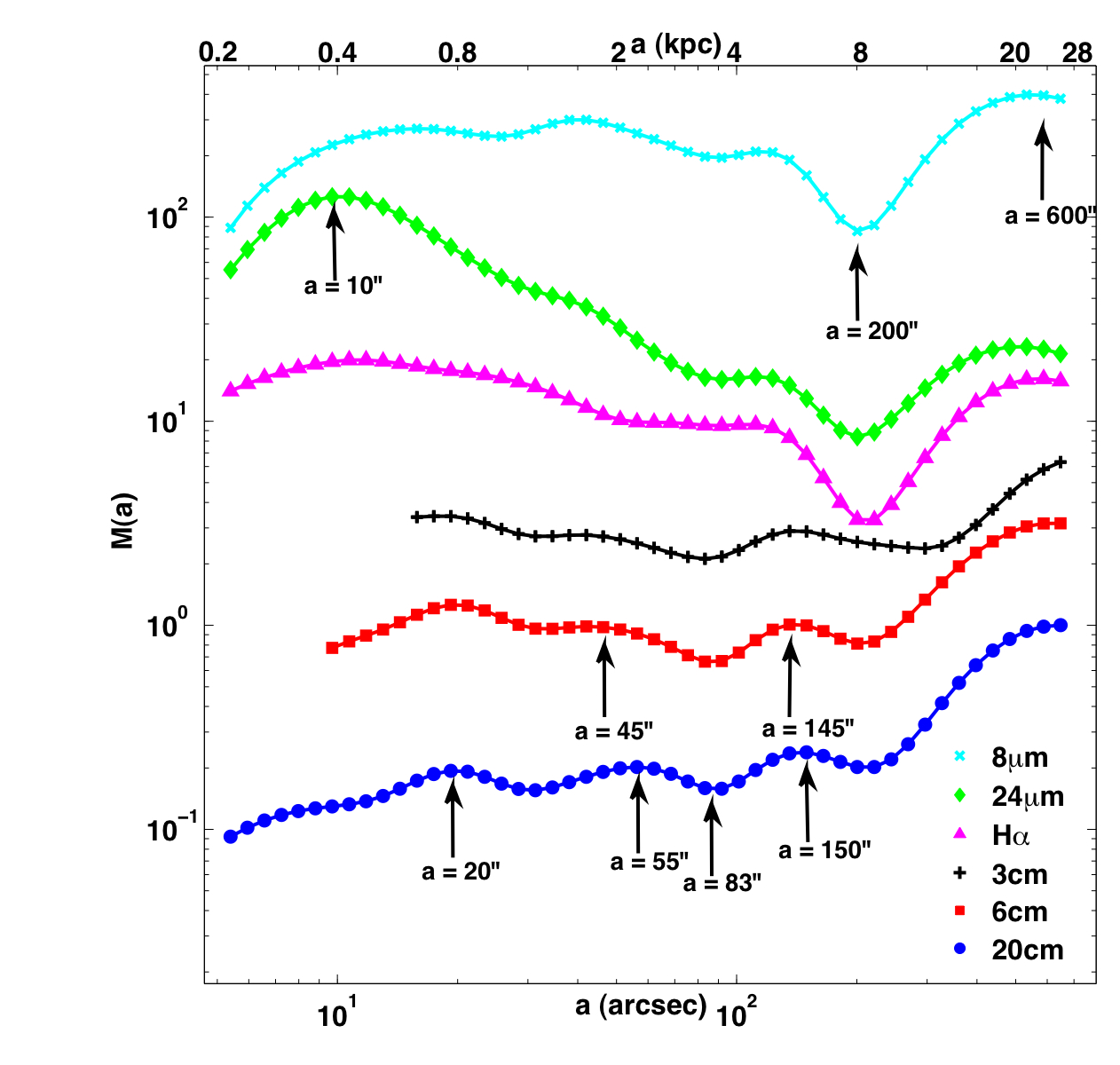} 
\caption{Wavelet spectra of the 20\,cm (blue circles), 6\,cm (red squares), 3.6\,cm (black pluses), H$\alpha$ (pink triangles), 24\,$\mu$m (green diamonds) and 8\,$\mu$m (cyan crosses) images at 2.4\arcsec\ resolution. The spectra are shown in arbitrary units. The scales of the x-axis are presented in arcsec (bottom) and  kpc (top). The arrows indicate the positions of some interesting scales, discussed in the Sections~\ref{sec:wavespectra} and \ref{scales}}
\label{wave_spec}
\end{center}
\end{figure}

\begin{figure}
\begin{center}
\includegraphics[width=\textwidth]{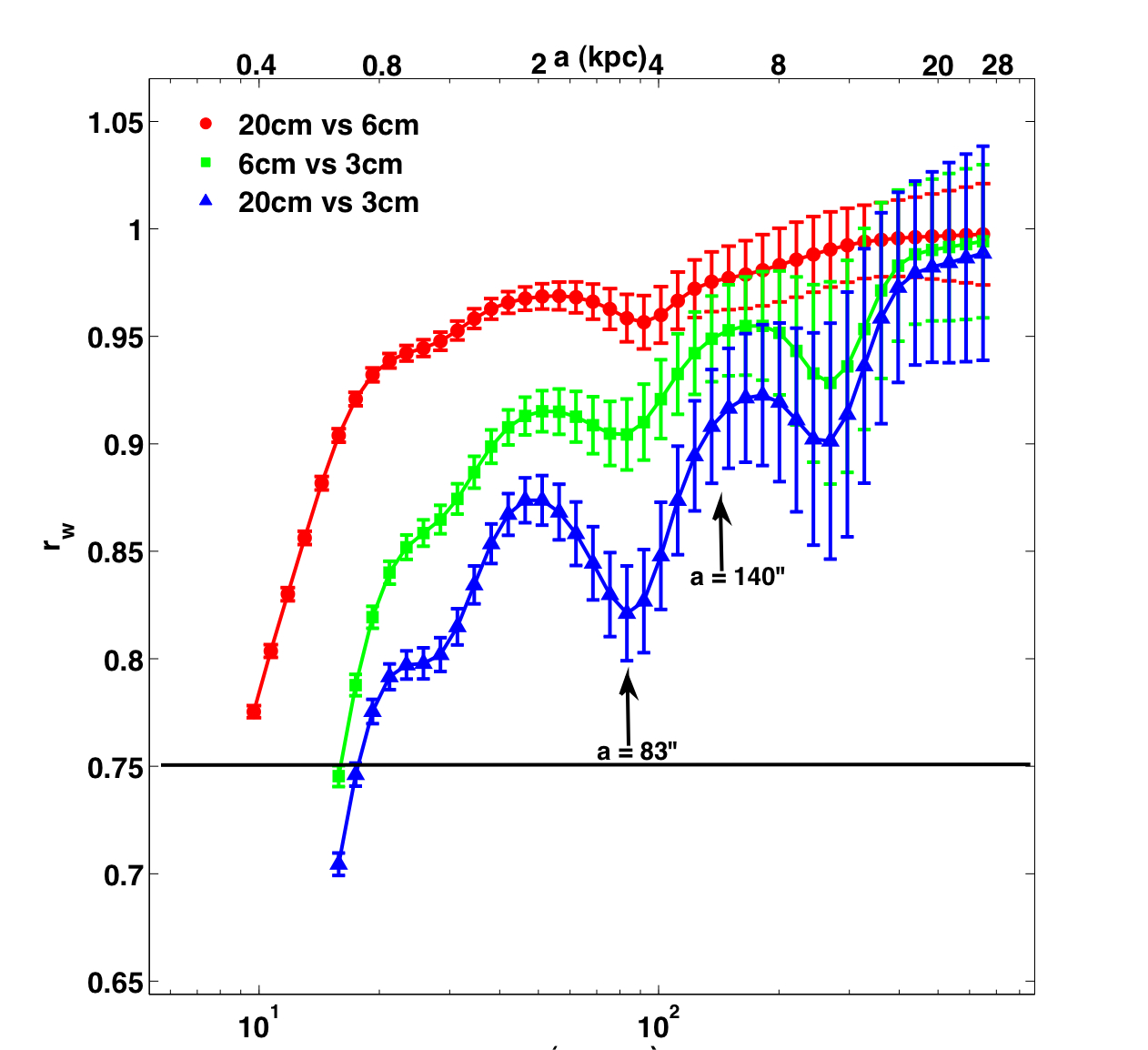} 
\caption{Wavelet correlation spectra between images at 20\,cm and 6\,cm (red circles), 20\,cm and 3.6\,cm (blue triangles) and 6\,cm and 3.6\,cm (green squares). The error bars have been computed following \cite{frick_01} (see Appendix~\ref{sec:wave_corr}). The horizontal line corresponds to a correlation of $r_w=0.75$, assumed to define significant correlation above this line (see text for details). The arrows indicate the positions of some interesting scales, discussed in Section~\ref{sec:crosscorr} and \ref{scales}}
\label{cross_corr_radio}
\end{center}
\end{figure}

\begin{figure}
\begin{center}
\includegraphics[width=\textwidth]{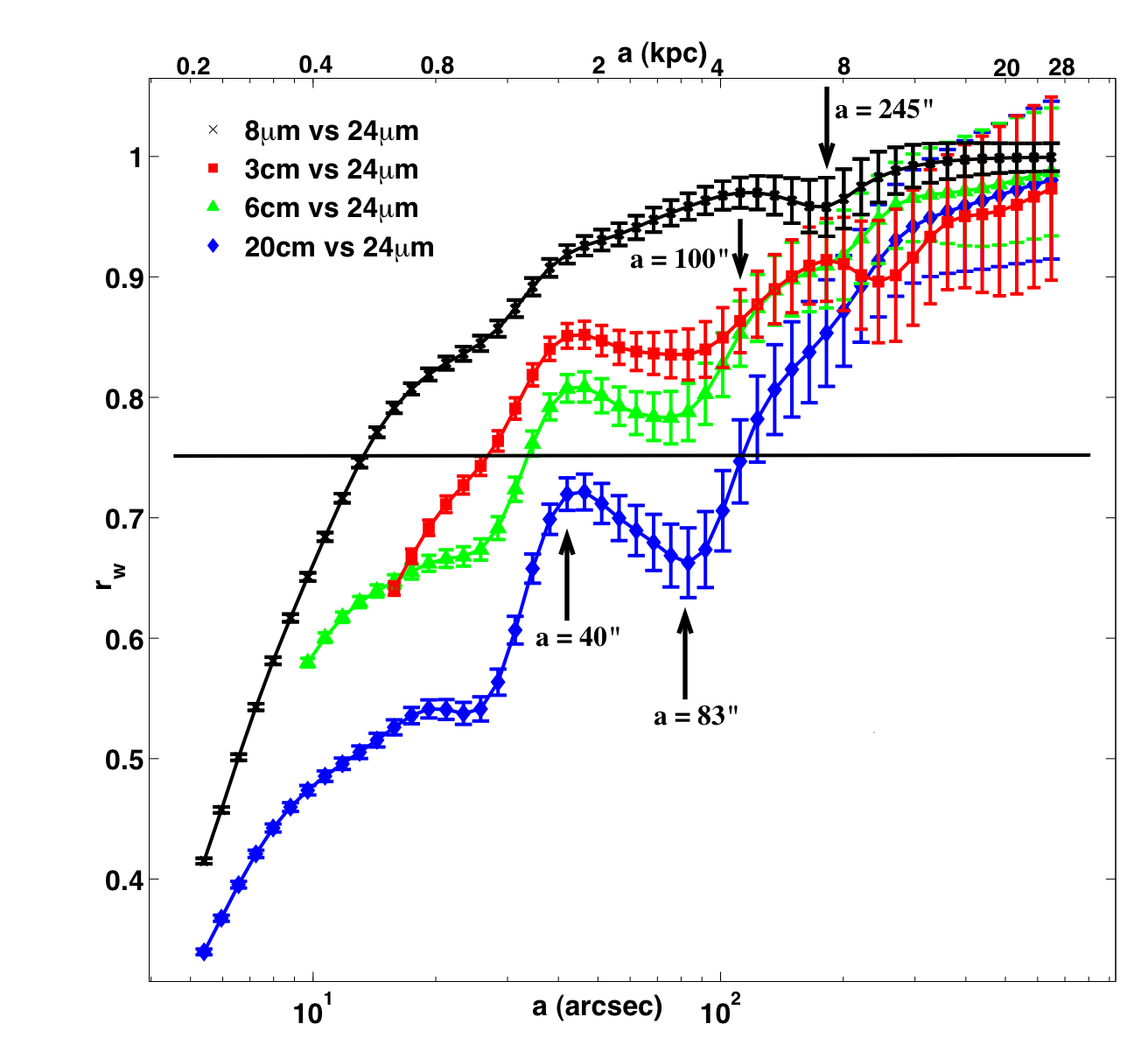}
\caption{Wavelet correlation spectra between 20\,cm and 24\,$\mu$m (blue diamonds), 6\,cm and 24\,$\mu$m (green triangles), 3.6\,cm and 24\,$\mu$m (red squares) and 8\,$\mu$m and 24\,$\mu$m (black crosses). The arrows indicate the positions of some interesting scales, discussed in Section~\ref{sec:crosscorr} and \ref{scales}}
\label{cross_corr_IR}
\end{center}
\end{figure}

\begin{figure}
\begin{center}
\includegraphics[width=\textwidth]{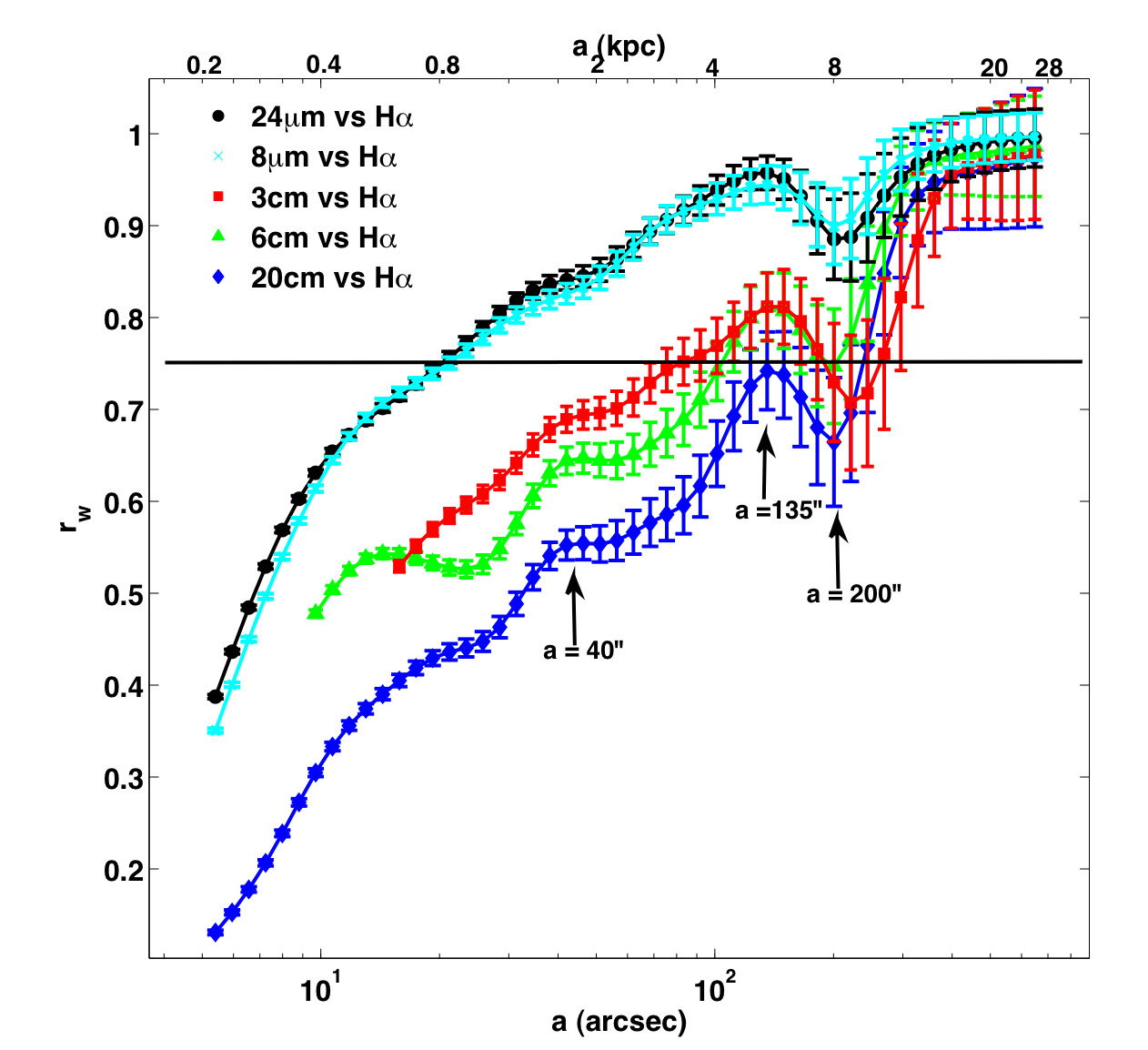}
\caption{Wavelet correlation spectra between 20\,cm and H$\alpha$ (blue diamonds), 6\,cm and H$\alpha$ (green triangles), 3.6\,cm and H$\alpha$ (red squares), 24\,$\mu$m and H$\alpha$ (black circles) and 8\,$\mu$m and H$\alpha$ (cyan crosses).  The arrows indicate the positions of some interesting scales, discussed in Section~\ref{sec:crosscorr}}
\label{cross_corr_ha}
\end{center}
\end{figure}

\begin{figure}
\begin{center}
\includegraphics[width=15cm]{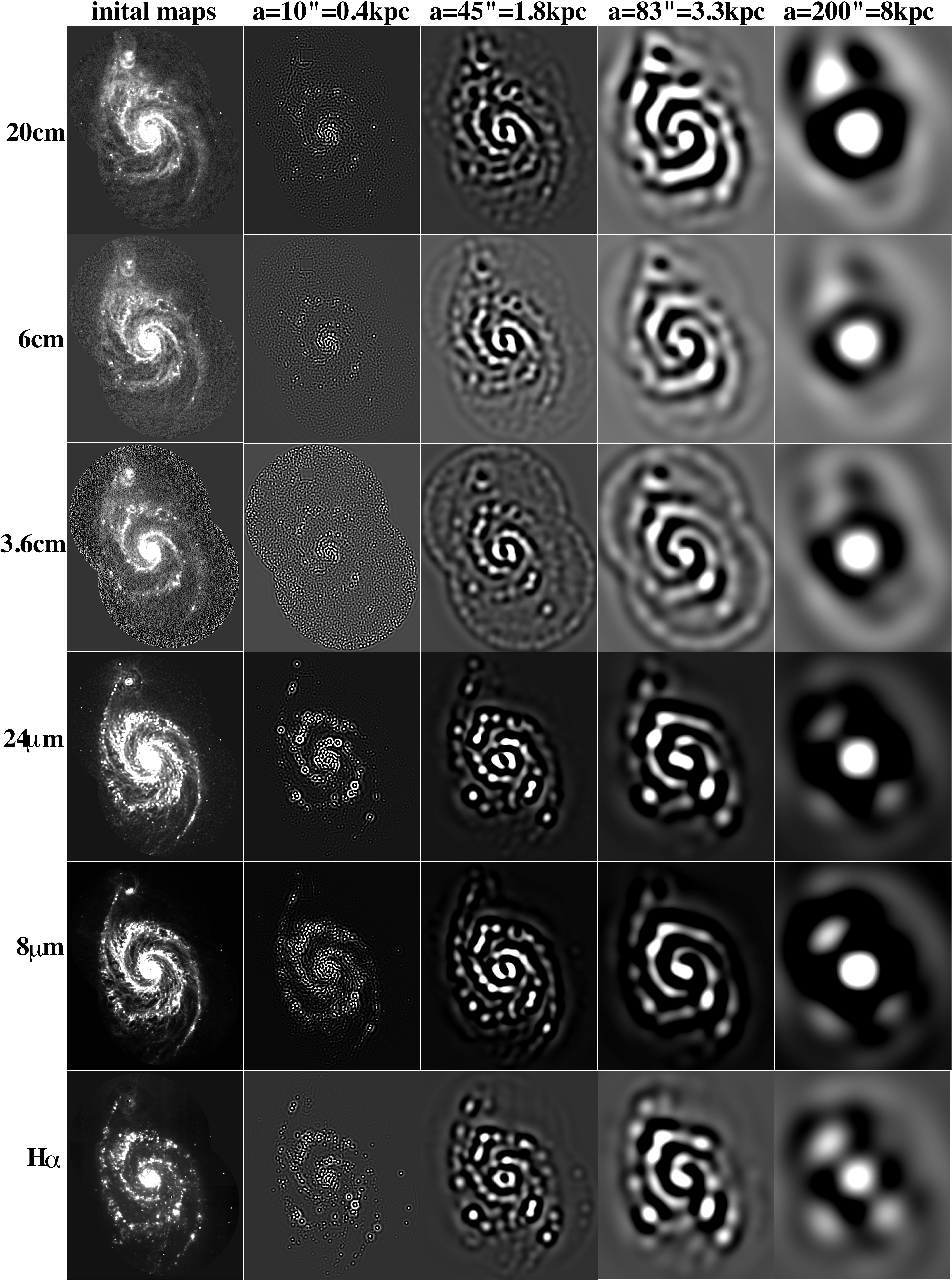}
\caption{Original maps (left) of the 20\,cm, 6\,cm, 3.6\,cm, 24\,$\mu$m, 8\,$\mu$m and H$\alpha$ emissions of M51 (from top to bottom) and the corresponding wavelet decomposition maps at four selected spatial scales. The field of view is 9\farcm 4$\times$11\farcm 2.}
\label{wave_transform}
\end{center}
\end{figure}

Here, we discuss important spatial scales in M51a galactic disk, as revealed by the wavelet spectra and cross-correlations, and their physical implications.
Fig.~\ref{wave_transform} shows the original maps smoothed to 2\farcs 4 resolution at 20\,cm, 6\,cm, 3.6\,cm, 24\,$\mu$m, 8\,$\mu$m and H$\alpha$ from top to bottom, and the corresponding wavelet decomposition maps at 4 prominent spatial scales: 10\arcsec\ (450\,pc), 45\arcsec\ (1.8\,kpc), 83\arcsec\ (3.3\,kpc) and 200\arcsec\ (8\,kpc). The wavelet decomposed maps show the structures which contribute at the different scales. 
The wavelet maps at 10\arcsec\ (Fig.~\ref{wave_transform} second column) show small structures which are dominating the wavelet spectrum at 24\,$\mu$m (see Fig.~\ref{wave_spec}). Those features also contribute largely to the 8\,$\mu$m and H$\alpha$ emission, although their spectra do not show a pronounced peak at this scale. They are formed of compact structures (especially in H$\alpha$) corresponding to \HII\ regions and filament-like features (especially at 8\,$\mu$m).

 At 45\arcsec\ ($\sim$1.8kpc)  we distinguish elongated structures defining the spiral arms (Fig.~\ref{wave_transform}, third column).  This spatial scale  is of the order of the typical width of the spiral arms of M51a. At this spatial scale the cross-correlations between the radio/IR/optical data reach a local maximum (Sect.~\ref{sec:crosscorr}). It is interesting to note that some filamentary structures in the inter-arm regions  contribute to the energy at this spatial scale, the length of these filaments being of the same order of the width of the spiral arms.

 Finally, at 200\arcsec (8\,kpc) ,  the wavelet decompositions show the diffuse emission  from the galactic disk and the region between M51 and its companion (north-east). This interaction zone dominates at long radio wavelengths (20\,cm and 6\,cm), it is weaker at 3.6\,cm and disappears in the IR and optical maps.  This spectral trend indicates  that the emission in this region is  mainly due to non-thermal radiation, such as synchrotron emission. The spatial scale corresponds also to a local minimum in the radio correlations between H$\alpha$ emission and the radio and IR emissions, likely due to this extra radio component in the interaction region (Fig.~\ref{cross_corr_ha}).

As shown in the previous sections, wavelet analysis is a powerful tool to determine the dominant morphological structures in our multi-wavelengths images. In the case of M51a, the scale at $a\,=\,83\arcsec\,\approx\,3.5\,$kpc is particularly interesting. It corresponds to a local minimum in the wavelet spectrum of all radio and IR images  (Fig.~\ref{wave_spec}),  which appears deeper at low than high radio frequency, as well as in the correlation between the radio and IR maps (see Sect.~\ref{sec:crosscorr}).  This characteristic is also seen in NGC6946 \citep{frick_01} and M33 \citep{fatemeh_07} at roughly the same spatial scale, 3.5\,kpc. We interpret this minimum due to the best contrast between the arm/interarm region. Indeed at this scale, the positive part of the wavelet function filters the structures formed by the combination of the bright spiral arms and the "darker" interarm regions. The resulting wavelet coefficients contain then information from these two regions and on the spatial scale where the contrast between spiral arms and interam is optimal, the wavelet coefficients are then lower than at smaller or larger scales, for which the contrast is less good and then more energy is contained into the positive part of the wavelet function. This leads to a local minimum in the wavelet spectrum at the corresponding spatial scale. Thus the wavelet transforms at the scale of $83\arcsec$  of M51a's radio and IR images trace well the spiral arm structure (see Fig.~\ref{wave_transform}, fourth column). We will use this result to investigate the local radio-IR correlation in M51a in the  following  two sections. First we discuss the wavelet cross-correlation between the 20\,cm and 24\,$\mu$m maps, then we use the wavelet transform at $83\arcsec$ to define different regions in M51a's galactic disk (spiral arms; interarm region; outer and inner regions) and discuss the correlation between 20\,cm and 24\,$\mu$m as a function of the environment.

\subsection{Spiral arms at 20\,cm and 24\,$\mu$m}

The local radio-IR correlation is formed on one hand by the strong relation between the thermal radio (Bremstrahlung) and the warm dust emission present in  \HII\ regions, and by the, albeit weaker, correlation between the synchrotron emission and the cool dust \citep{hoernes_98}. Here we discuss the correlation between the 20\,cm radio continuum, which is the most sensitive tracer for synchrotron radiation in our radio data \citep{condon_92} and the 24\,$\mu$m emission within M51a.  We use the 24$\mu$m emission to investigate the radio-IR correlation since it has the same spatial resolution as our radio maps, although it is mostly related to warm dust. 
Synchrotron and IR continuum exhibit many complex structures over all spatial scales as shown in Figures\,\ref{fig:HIRES} and \ref{fig:radio}. Moreover they do not always arise from the same location, and in some regions, non-thermal radio emission is not at all associated with IR  emission, as in the interaction region between M51a and NGC5195. 
CR electrons are usually produced and accelerated in  SNRs and radiate synchrotron emission while propagating in the magnetized interstellar medium. In the regions where radio emission is not spatially associated to IR, synchrotron emission from CR electrons may be enhanced (e.g. by strong magnetic field) or produced by a different physical process (such as galactic-scale shocks) rather that SF induced shocks. 

 The wavelet spectra reveal that the predominant structures at 20\,cm and 24\,$\mu$m are different (see Sec.~\ref{sec:wavespectra}). While small scale structures dominate the IR emission, they carry  only little energy in the radio continuum maps (Fig.~\ref{wave_spec}). Those structures correspond to bright knots in the spiral arms, associated with \HII\ regions and stellar clusters (Fig.~\ref{wave_transform}).  The spiral arms at 20\,cm are wider than at the other wavelengths ($\sim\,60\arcsec$ compared to about 40\arcsec\ in 3.6\,cm and IR maps as traced by the maximum in the wavelet spectra, see Sect.~\ref{sec:analysis} and Table~\ref{wave_minmax}), reflecting the diffuse synchrotron emission extending into the interarm regions.  On larger scales (83\arcsec\ $\approx$ 3.5kpc, see Fig.~\ref{cross_corr_IR})  the correlation between 20\,cm and 24\,$\mu$m emission reaches a local minimum.  The two wavelet transform maps at this spatial scales show similar structures, tracing the spiral arms and interarm region (Fig~\ref{wave_transform},  fourth column). However, an obvious difference between these two maps is that the spiral arm structure revealed by the wavelet transform of the 20\,cm emission at this spatial scale is more continuous than the structures tracing the arms in the 24\,$\mu$m wavelet transform. This could be explained by the diffusion of the CR electrons from their acceleration site into the ISM.

 The spatial scale on which the correlation between the NT-radio continuum and the IR emission should break is generally assumed to be the diffusion length of CR electrons. Following our wavelet analysis, the correlation between 20\,cm and 24\,$\mu$m starts to decrease rapidly from scales of 45\arcsec\ (1.8\,kpc) downwards. Between 30\arcsec\ and 20\arcsec\ (1.2-0.8\,kpc) the correlation coefficients get stabilized around a value of 0.54 and then decrease further from  0.5 - 0.8 kpc towards smaller spatial scales,  in good agreement with CR electrons' diffusion lengths of about 0.5\,kpc in M51 estimated by \cite{murphy_06,murphy_2008}. This shows that the radio-IR correlation is not smooth and exhibiting local maxima and minima at different spatial scales corresponding to particular morphological structures in the galaxy. This result reflects the complexity of the local radio-IR correlation and hints that different processes lie at the origin of this relation 
  and that while one process may dominate at a particular spatial scales, none is dominating at all scales.  In particular, the fraction of non-thermal and thermal emission can change with spatial scales, while diffuse IR emission form older stellar population can contaminate the emission at 24\,$\mu$m considered here. These different components of the radio and IR relation can explain the observed variations in the radio/IR relation (traced here by the correlation between the 20cm and 24\,$\mu$m emission). This will be developed in details in a further study, where in particular careful decomposition of the radio continuum into thermal and non-thermal components will be done (Dumas et al., in preparation). The decrease of the correlation coefficients towards smaller spatial scales can be interpreted as a break of correlation between the radio and IR emission on those scales.   In particular it is interesting to note that the correlation between the 20cm and 24$\mu$m emission does not hold on typical scales of star forming complexes. The break of correlation on these small scales is expected given the timescales difference between the IR and non-thermal radio emission and the diffusion length of cosmic rays in the ISM.

\subsection{Observed environmental effects}

\begin{figure}[h!]
\begin{center}
\includegraphics[width=\textwidth]{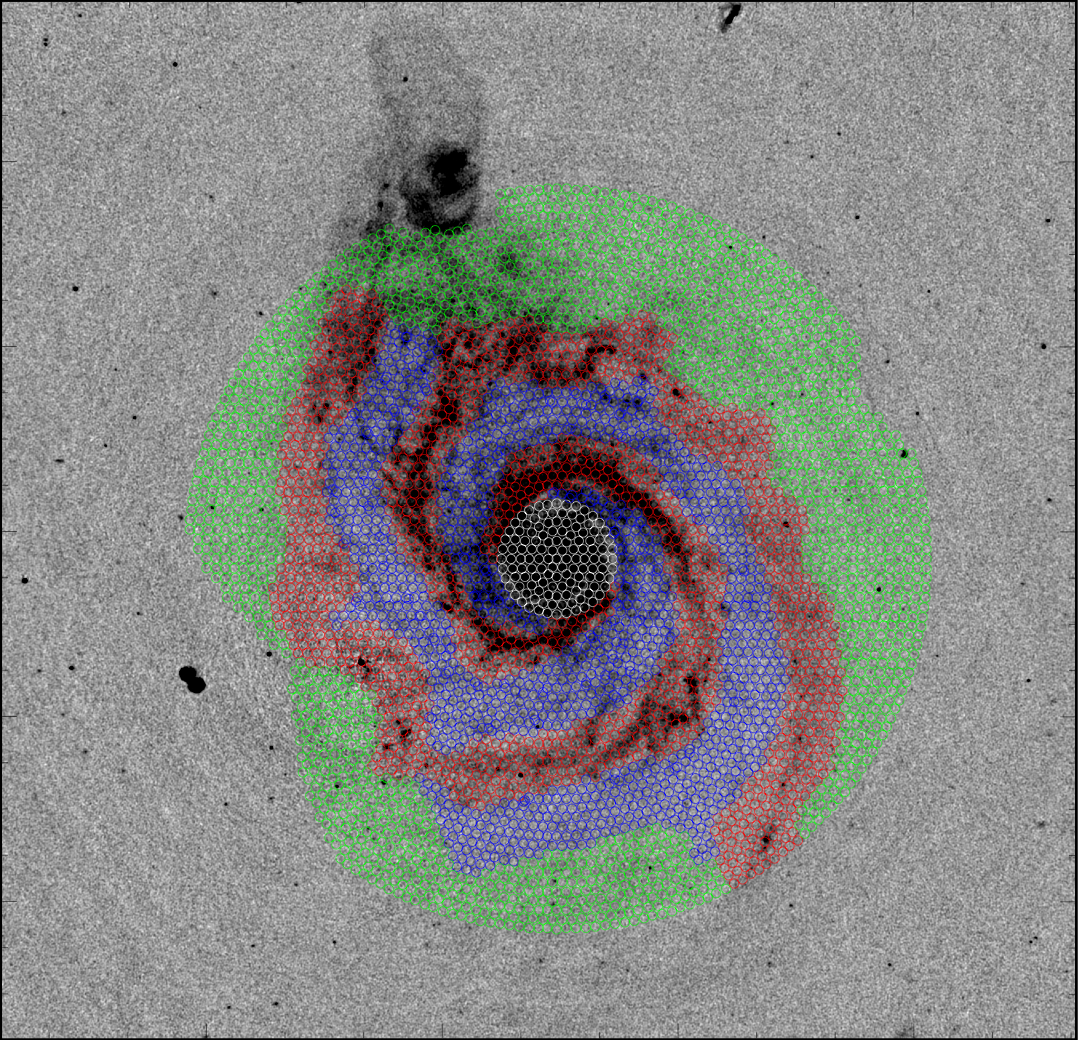} 
\caption{6\arcsec\ diameter apertures plotted on the 20\,cm image. The apertures fill the different regions studied: the central region (white), the spiral arms (red), the inter-arm region (blue) and the outer disk (green).}
\label{apertures}
\end{center}
\end{figure}

\begin{figure}[h!]
\begin{center}
\includegraphics[width=8cm]{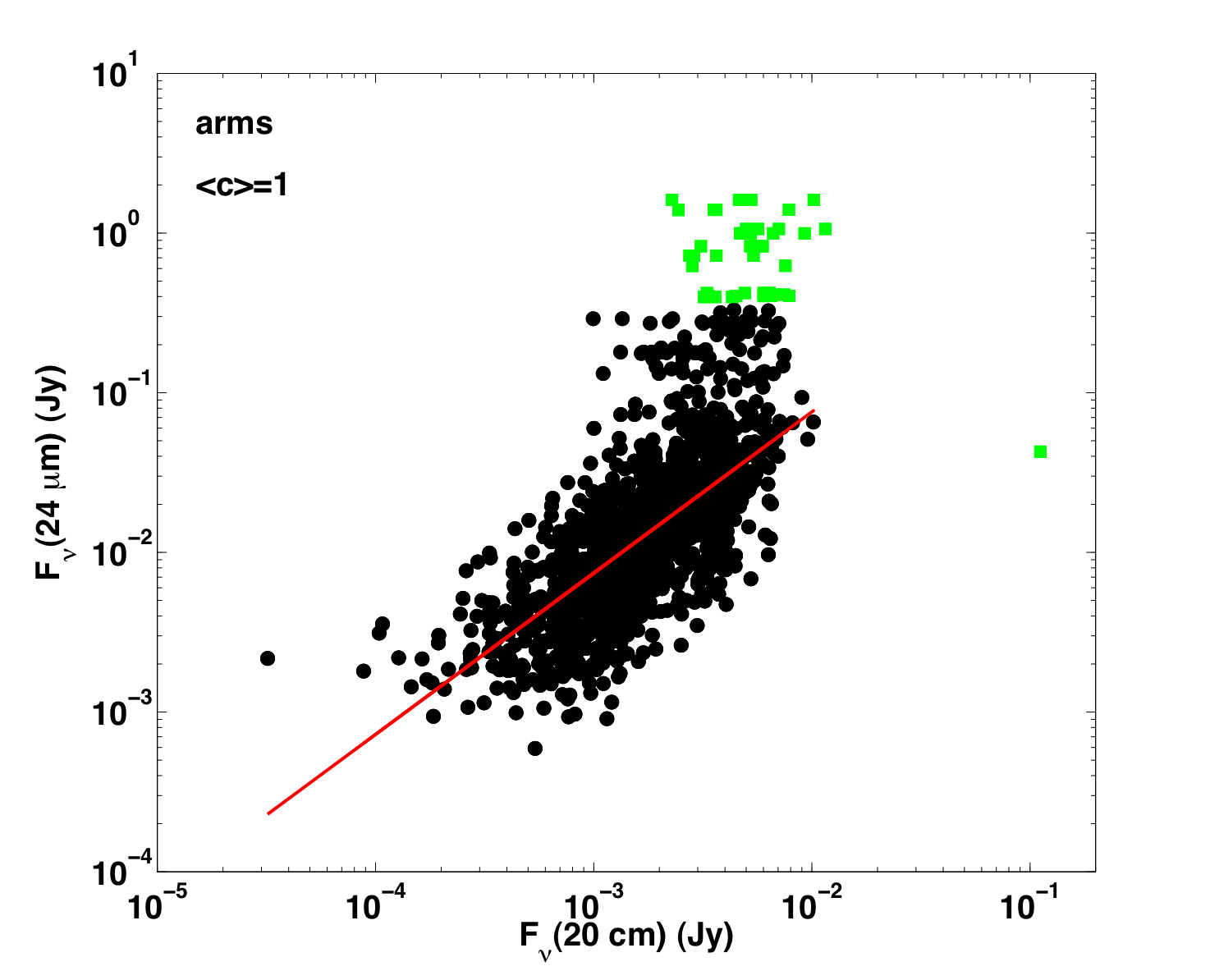} 
\includegraphics[width=8cm]{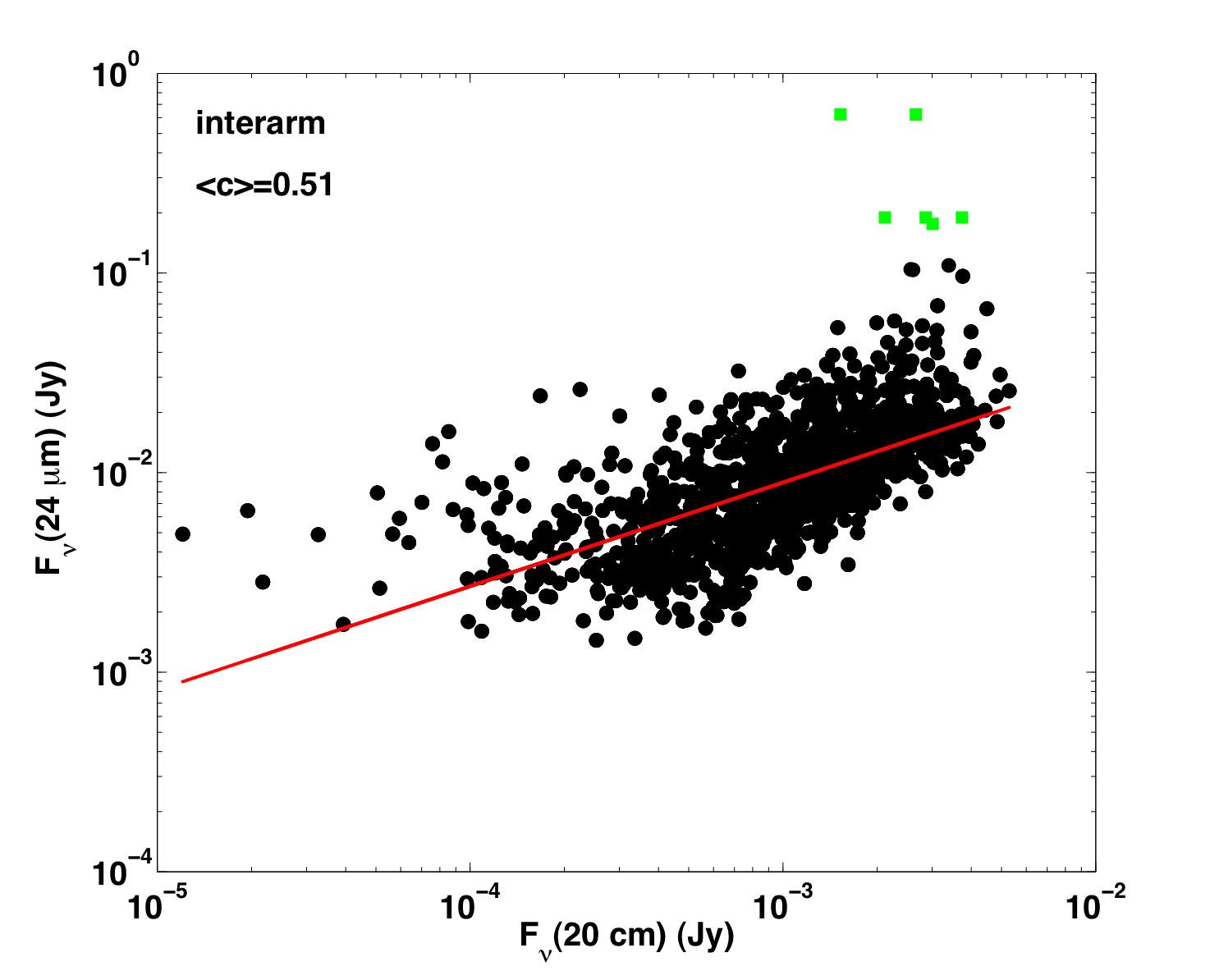} 
\includegraphics[width=8cm]{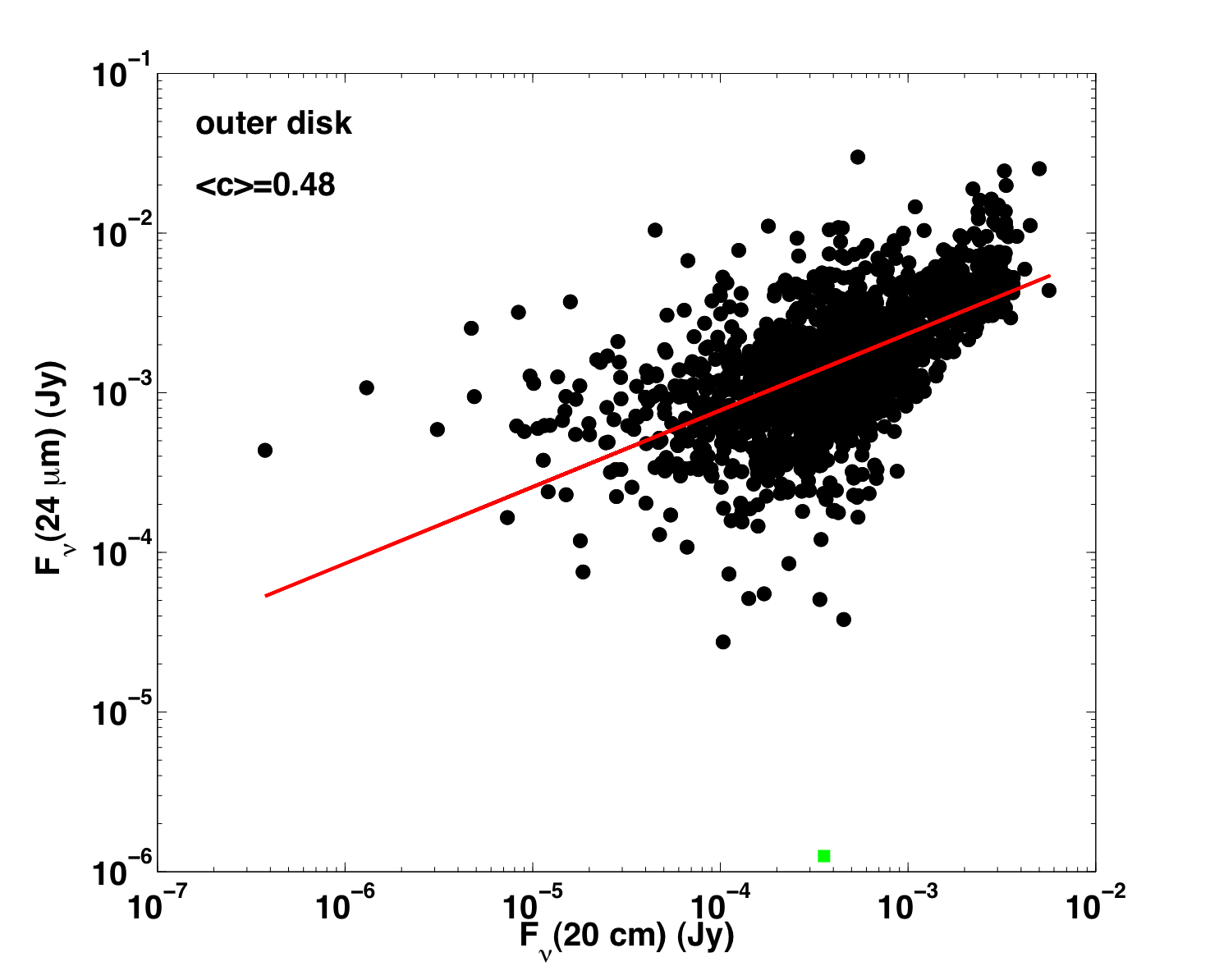} 
\includegraphics[width=8cm]{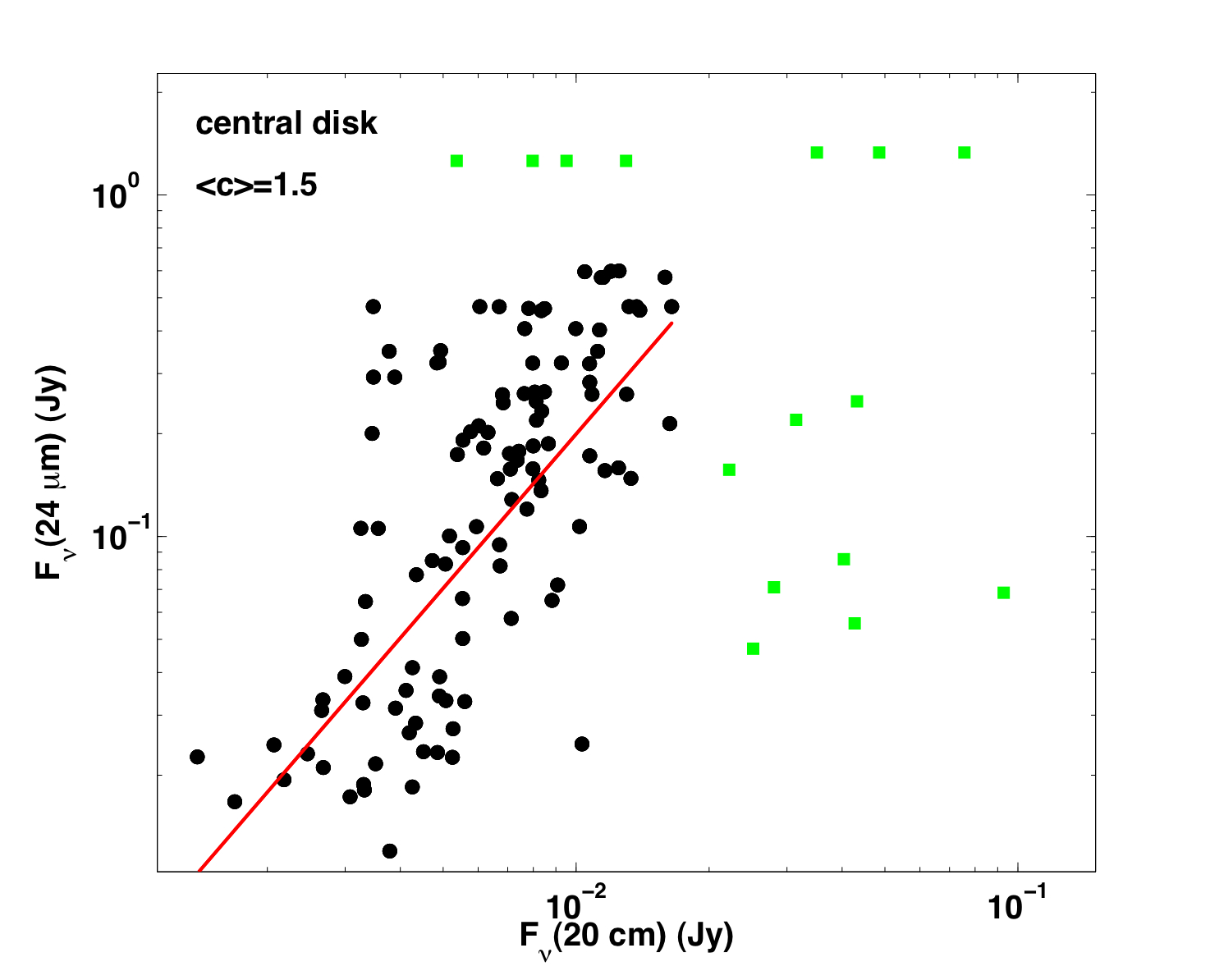} 
\caption{24\,$\mu$m flux density plotted against 20\, cm flux density computed within 6\arcsec\ diameter apertures for four regions defined in the galactic disk (see text): spiral arms (top left), interarm region (top right), outer region (bottom left) and inner region (bottom right).  The red  lines correspond  to the fit derived here (see Tab.~\ref{fit_values}). Outliers excluded from the fits are marked in green.}
\label{24micron_vs_20cm}
\end{center}
\end{figure}

In order to investigate the radio-IR correlation in the different environments of M51a, we identified different regions within the galactic disk. The spiral arms and interarm region  are defined by the wavelet transform at the spatial scale $a\,=\,83\arcsec$, as this spatial scale provides the best contrast between the arm/interarm region (see Sec.~\ref{sec:wavespectra}). Positive values of the wavelet filtered 20\,cm image at 83\arcsec\ define the spiral arms, while negative values mask the interarm region. The region with R$\,<\,$1.75\,kpc is the inner region and the outer disk region starts at the outer edges  of the external spiral arms, at roughly R$\,>\,6.5$ kpc. We excluded the companion NGC5195 from the analysis. The different regions are filled with apertures of 6\arcsec\ diameter. Figure~\ref{apertures}  shows the different regions overlaid on the 20\,cm  map. The 20\,cm and 24\,$\mu$m flux densities within these apertures are calculated with the task 'aphot' in IRAF. 
Fig.~\ref{24micron_vs_20cm} presents the flux densities at 24\,$\mu$m and 20\,cm  in the spiral arms (top left panel), interarm (top right panel), outer disk (bottom left panel) and inner region (bottom right panel).

\begin{figure}[h!]
\begin{center}
\includegraphics[width=\textwidth]{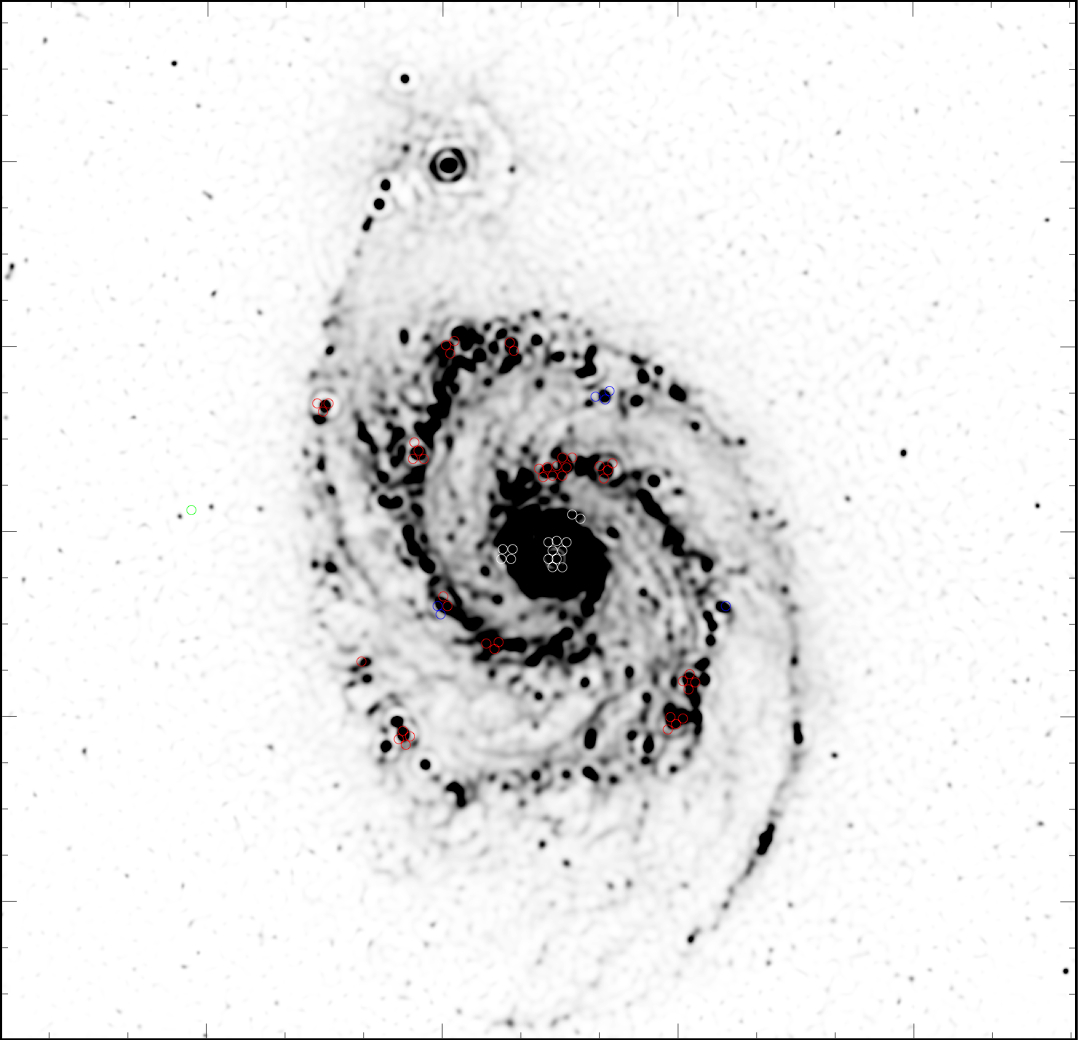} 
\caption{Positions of the outlier points which have been removed prior to the fit between 24\,$\mu$m and 20\,cm fluxes in the inner region (white circles), spiral arms (red circles), interarm region (blue circles) and outer disk (green circle).}
\label{outliers}
\end{center}
\end{figure}

 We fitted the relation between the 24\,$\mu$m and 20\,cm emission with the function $F_{24\,\mu m} = b \times (S_{20\,cm})^c$, with $F_{24\,\mu m}$ and $S_{20\,cm}$ being the flux density at 24$\,\mu$m and at 20\,cm (in Jy), respectively.   The fitted values and standard deviations of the scaling exponent $c$ and the coefficient $b$, are listed in Table.~\ref{fit_values}. Prior to this fit, some points in the different regions were excluded. Those points are marked in Fig.~\ref{outliers}. They correspond to AGN related emission (in particular the radio jet) and bright \HII\ regions in the inner region (white circles in Fig.~\ref{outliers}), to bright \HII\ regions in the spiral arms and interarm region (respectively, red and blue circles in Fig.~\ref{outliers}) and to one point at very low 24\,$\mu$m flux density in the outer region (green circle in Fig.~\ref{outliers}).  Those outliers are also highlighted as green squares on Fig.~\ref{24micron_vs_20cm}. The results of the fit in each region are given in Table~\ref{fit_values} and plotted as red lines in Fig.~\ref{24micron_vs_20cm}.\\
 The relation between $F_{24\,\mu m}$ and $S_{20\,cm}$ is linear in the spiral arms and globally over the galaxy. In this case,  the logarithm of parameter $b$ is equal to the parameter $q_{24}$  defined by \cite{murphy_06} as $q_{24}\,\equiv\,\log\left[ \dfrac{F_{24\,\mu m} (Jy) }{S_{20\,cm} (Jy)}\right]$ and we find that our fitted parameter is in good agreement with the values derived by \cite{murphy_06} (see Table~\ref{fit_values}). In the interarm region and outer disk, the flux density at 24\,$\mu$m is proportional to about the square-root of the 20\,cm flux density while in the inner regions we observe an opposite behavior: $F_{24\,\mu m} \propto  (S_{20\,cm})^{1.5}$. In these regions where  a non-linear relation occurs between $F_{24\,\mu m}$ and $S_{20\,cm}$, $q_{24}$ is determined by both $\log(b)$ and $c$, hence is not constant and of a limited use as a quantitative measure of the radio-IR correlation. 

The large differences in the exponent $<c>$ of the infrared-radio
correlation (Table~\ref{fit_values}) demonstrate that the physical
processes involved are different in the arm, interarm and inner regions,
while the interarm and outer regions show a similar behaviour. As the spiral arms
dominate the total flux emission (50\% at 20\,cm and 60\% at 24\,\micron), a linear behaviour is obtained when looking at the whole galaxy, as observed for the
integrated emission of most galaxies \citep{yun_01,bell_03}. Our results are
especially striking in view of the many processes involved: star
formation, dust heating, dust opacity, cosmic-ray acceleration,
cosmic-ray propagation and magnetic fields \citep[see e.g.][]{bicay_90, murphy_06}. In spite of the complicated
interplay of the ISM's different components, three well-defined regimes can be
distinguished and are discussed in detail below.

\subsection{Possible causes for environmental dependencies}

Here, we are following the formalism and assumptions used by \cite{niklas_97}. 
Assuming equipartition between the energy densities of the magnetic field and CR electrons, we can couple the radio non-thermal flux density to the gas volume density. The energy density equipartition assumes that the energy densities of cosmic rays (electrons and protons) and magnetic field are approximately equal and gives:

$$
S_{20\,cm}\,\propto\,B_{eq}^{3+\alpha} 
$$
where   $B_{eq}$ is the total equipartition magnetic field strength
\citep{beck_05} and $\alpha$ is the synchrotron spectral index.  The equipartition magnetic field strength $B_{eq}$ is also referred to as the minimum energy magnetic field as it is the minimum value of $B$ to realize this energy equipartition. This assumption has been and is still discussed \citep{niklas_97, beck_05, thompson_06}. In particular \cite{thompson_06} compare this value of the magnetic filed strength measured in different galaxies to different scaling relations between $B$ and the gas surface density and show that the equipartition magnetic field strength underestimates the magnetic field in starburst galaxies. M51 however follows the relation of normal star forming galaxies for which the magnetic energy density of the equipartition magnetic field  is comparable to the total hydrostatic pressure of the ISM \citep[see Eq.~3 and Fig.~1 of][]{thompson_06}. In that case we can assume that the equipartition magnetic field strength reflect to total magnetic field in M51 and that energy density equipartition is a good assumption for this galaxy.
In M51 \cite{fletcher_10} find $\alpha\,\simeq\,
0.6$ in the spiral arms and inner region, and $\alpha\,\simeq\,1.2$ in
the interarm and outer regions. In addition, the magnetic field is coupled to the gas:

$$
B\,\propto\,\rho_g^{\beta},
$$
 $\beta$ describes the coupling between the magnetic field and the gas volume density $\rho_g$. For
turbulent field amplification (e.g. by dynamo action),
$\beta\,\simeq\,0.5$ is expected \citep{beck_96, thompson_06}. This value also
 emerges from this time equipartition between magnetic and turbulent
energy densities. A similar value is predicted for cloud collapse
along magnetic field lines \citep{mouschovias_76}, while isotropic collapse in turbulent
fields gives $\beta\,\simeq\,2/3$. In case of one-dimensional
compression or shear of magnetic fields by large-scale flows of
diffuse gas, $\beta\,\simeq\,1$ is expected.

We therefore obtain a relation between the radio NT emission and the gas volume density:
$$S_{20\,cm}\,\propto\,\rho_g^{\beta(3+\alpha)}$$

On the other hand we can also link the 24\,$\mu$m flux density to the gas density. In the case of optically thick
dust for UV photons, the dust re-processes the totality of the radiation of the ionizing stars and thus the IR emission traces the star formation rate (SFR). 
 Therefore we assume: 
$ F_{24\,\mu m}\,\propto\,I_0  $ where $I_0$ is the energy density of the dust-heating radiation field and $I_0\,\propto\,SFR$ 
where $SFR$ is the star-formation rate, hence $F_{24\mu m} \propto SFR$. 
Moreover the Schmidt law  \citep{schmidt_59} gives a relation between the star formation rate SFR and the gas volume density $\rho_g$:   $SFR \,\propto\,\rho_g^{n} $, with $n$ being the Schmidt law index. Therefore we can write:
$$ F_{24\,\mu m}\,\propto\,\rho_g^{n}  $$
 and  $n=1.4 \pm 0.3 $ \citep{niklas_97}. 
In that case, the correlation between $F_{24\,\mu m}$ and $S_{20\,cm}$ with the exponent $c$ yields
\begin{equation}
\label{eq:opt_thick}
\beta\,=\,\dfrac{n}{c(3+\alpha)}
\end{equation}

If the dust is not optically thick for UV photons, as, e.g., in the
galaxy M31 \citep{hoernes_98}, then the optical depth of the dust  $\tau$ has to been taken into account:
$$
F_{24\,\mu m} \propto \tau \, I_0
$$
 $I_0$ is the energy density of the dust-heating radiation field directly proportional to the star formation rate, and $\tau\,\propto\,\rho_{dust}$, then assuming a constant dust-to-gas ratio we have a relation between the 24\,$\mu$m IR flux and the gas: 
$$F_{24\,\mu m} \,\propto\,\rho_g^{(1+n)}
$$
 
and Eq.~\ref{eq:opt_thick} becomes
\begin{equation}
\label{eq:opt_thin}
\beta = \dfrac{n+1}{c(3+\alpha)}
\end{equation}

As shown in \cite{sings_2005}, in most regions of M51, (e.g. the spiral arms)  dust attenuation is high enough that the IR emission traces well the SFR and Eq.~\ref{eq:opt_thick} can be applied. However, in region where dust extinction is lower and diffuse IR emission, coming from dust heated by the old stellar component (the so-called 'Cirrus' emission) dominates, as at large radii or in the interarm region, IR emission is not totally related to recent star formation.  The IR emission for a fixed SFR will be lower in such regions. 
  And in the case of diffuse Cirrus emission we have the following relations: 
$F_{cirrus}\,\propto\,\rho_{dust}\,\propto\,\rho_g$ , where $ \rho_{dust}$ and $ \rho_g$  are the volume density
of  dust and gas respectively (we assume a constant dust-to-gas ratio).  Then, $F_{24\,\mu m}\,\propto\,\rho_g$ and Eq.~\ref{eq:opt_thick} becomes
\begin{equation}
\label{eq:cold_dust}
\beta = \dfrac{1}{a(3+\alpha)}
\end{equation}

We have derived three possible relations between the IR and 20cm emission, depending on the properties of the ISM. We now discuss the three trends found  between the radio and IR emission in the spiral arms, interarm region, outer disk and inner region of M51a: 

\begin{enumerate}
\item In the spiral arms, we can assume the dust to be optically thick and directly apply 
Eq.~\ref{eq:opt_thick}. Therefore, with $<c>\,=\,1.01\,\pm\,0.01$, $\alpha\,=\,0.6\,\pm\,0.02$ \citep{fletcher_10} and
$n = 1.4\,\pm\,0.3$, we obtain $\beta = 0.4\,\pm\,0.1$ which is slightly  below the
range of theoretical predictions, but with an overall good agreement. The expected value  under the assumption that magnetic density energy is comparable to the turbulent energy density of the ISM is $\beta\,=\,0.5$ \citep{beck_96, thompson_06}. Fig.~1 of \cite{thompson_06} shows that the equipartition magnetic field strength in M51 realizes this assumption., and in that case, the Schmidt law index should be $n\,=\,1.8\,\pm\,0.2$. 

\item In the interarm and outer region,  the spectral index is higher $\alpha\,=\,1.2\,\pm\,0.06$ \citep{fletcher_10}. Therefore,  with $<c>\,=\,0.51\,\pm\,0.01$,
 and $n\,=\,1.4\,\pm\,0.3$, we obtain
$\beta\,=\,0.7\,\pm\,0.1$ in the case of optically thick dust.  Assuming UV-thin dust for the interarm and outer regions, we obtain $\beta\,=\,1.0\,\pm\,0.1$, 
as expected for compression or shear of magnetic fields in flows of diffuse gas. This possibility should be further investigated by comparing high-resolution data 
in radio continuum, \HI\ and CO.

 In addtion, the Cirrus IR emission may provide another interpretation of the non-linearity between $F_{24\,\mu m}$ and $S_{20\,cm}$
in the interarm and outer regions. In this region, we find that the radio-IR correlation is sub-linear, which implies that for a fixed SFR the 24\,$\mu$m is lower that predicted (assuming 20\,cm traces the SFR), as expected for regions dominated by Cirrus IR emission. 
 Applying Eq.~\ref{eq:cold_dust} we then find $\beta\,\simeq 0.5$ in the interarm and outer regions, as in the spiral arms, which is in good agreement with theoretical values from turbulent field amplification \citep{beck_96}. Therefore, diffuse IR emission emitted by dust heated in large part by older stars seem a good explanation for the sub-linear radio-IR correlation  found in the interarm and outer regions.

Further high-resolution observations and measurements of dust opacity and separation of the cold and warm dust components are needed to distinguish between these two possibilities.

\item For the inner region we found $<c>\,=\,1.5\,\pm\,0.1$. In this region, the AGN related radiation, in particular the radio jet,
 has been excluded for the fit. With $\alpha\,
\simeq\,0.6$ \citep{fletcher_10} and $n \,=\, 1.4\,\pm\,0.3$ we obtain $\beta\,=\,0.26\,\pm\,0.05$, in the case of optically thick dust. 
This value of $\beta$ is well below the range of theoretical predictions. The
equipartition field is too weak compared to the gas density. Either
the process amplifying the magnetic field is less efficient in the
inner region, or the equipartition assumption yields field strengths
which are too low. The former case needs further investigation which
is beyond the scope of this paper. The latter case is known to be
valid for strong synchrotron or inverse Compton losses of the
cosmic-ray electrons \citep{beck_05}. As the magnetic field
and the radiation field have large energy densities in the inner
region of M51a, the radio synchrotron emission is probably
suppressed, hence the radio-derived SFR underestimate the true amount of star formation present in the inner 1.5\,kpc.
\end{enumerate}

In summary, we can explain the different exponents of the
infrared-radio correlation seen in the different galactic environments 
 by a change in the optical dust opacity and the relations between the magnetic field strength and the gas density and the presence of a 'Cirrus' contribution to the IR emission.
 The steep exponent detected in the inner region is probably the effect of suppression of radio
synchrotron emission due to energy losses of cosmic-ray electrons. Future high-resolution
observations of galactic disks with HERSCHEL and the EVLA (and later
with the SKA) will provide an excellent tool to study the
interactions of the ISM components in galaxies. \\

\section{Conclusion}
\label{sec:concl}

In this paper we presented high quality radio continuum data of the nearby galaxy system M51 at three frequencies, 1.4\,GHz (20\,cm), 4.9\,GHz (6\,cm) and 8.4\,GHz (3.6\,cm), at a resolution between 1.5\arcsec\ and 2.4\arcsec. Combining these radio data with deconvolved Spitzer IR images, we were able to investigate the radio-IR correlation within M51 on spatial scales down to about 200\,pc. The wavelet analysis of the radio and IR maps of M51 reveals that the radio-IR correlation is not uniform accros the galactic disk and exhibits a complex behavior due to the different galactic structures present in the galactic disk. In particular, the radio-IR correlation is high within the spiral arms while it becomes weaker for scales encompassing the spiral arms and interarm regions, indicating that another process unrelated to star formation might contribute to the radio emission, such as a local enhancement of the synchrotron emission by the magnetic field. 

Using our wavelet analysis to define different environments within the disk of M51  we found that the IR-radio correlation is globally linear over the galaxy and within the spiral arms, in apertures of  6\arcsec (240\,pc) diameter,  reflecting that both IR and radio emission are good SF tracers in regions where dust exctinction is high.  For the first time we could quantitatively show that the radio continuum-IR correlation is not
linear  in the interarm and outer regions, where the 24\,$\mu$m emission is proportional to the  square-root of the 20\,cm radio continuum emission, as well as in the central part of M51 where the radio-IR correlation is over-linear. This can be well explained by variation of the dust opacity and the predominance of a diffuse IR emission coming from the old stellar population in the outer and interam regions, while the validity of estimates of magnetic field strengths based on the equipartition assumption  are
reflected in the exponent of the correlation in the inner regions.


\acknowledgments
The authors thank Daniela Calzetti for providing the H$\alpha$ image.  We also are thankful to A. Fletcher, R. Kennicutt and U. Lisenfeld for very fruitful discussions and to the anonymous referee for useful comments and suggestions that helped to improve this paper.
This work is partly based on observations made with the Spitzer
Space Telescope, which is operated by NASA/JPL/Caltech, on observations with the 100-m telescope of the MPIfR 
(Max-Planck-Institut f\"ur Radioastronomie) at Effelsberg and on observations made with the NRAO Very Large Array.
The National Radio Astronomy Observatory is a facility of
the National Science Foundation operated under cooperative
agreement by Associated Universities, Inc. G. D. was supported by DFG grants SCH 53614-1 and SCH 53614-2 as part of SPP 1177.

\bibliographystyle{aj}
\bibliography{M51_bib}

\appendix

\section{Wavelet analysis: mathematical formalism}
\label{app:wave}

\subsection{The 2-D continuous wavelet transform}
The wavelet analysis of an image is a generalisation of the Fourier analysis, where the sinusoidal waves are replaced by a family of self-similar functions that depend on scale and location. These functions are generated by dilating and translating and rotating a mother function which is called the analysing wavelet. \cite{frick_01} showed that the wavelet analysis is well adapted for the  investigation of scale-dependent properties in astrophysical objects. For this study, we restrict our analysis to the use of isotropic wavelets, we assume here that the structures that we will probe with such analysis are axisymmetric, hence with no favored orientation. We then do not apply any rotation to the mother wavelet. The continuous 2-dimensional wavelet transform  can be written as:
\begin{equation}
W(a,\boldsymbol{x}) = \omega(a) \int_{-\infty}^{\infty}\int_{-\infty}^{\infty} f(\boldsymbol{x'}) \psi^{*}\left( \frac{\boldsymbol{x'}-\boldsymbol{x}}{a}\right) d\boldsymbol{x'}
\end{equation}

where $a$  is the scale size, $f$ is a 2-dimensional function (the image), $\psi^*$ is the complex conjugate of the analysing wavelet,  $\boldsymbol{x}$ and $\boldsymbol{x'}$ are coordinate vectors and $\omega(a)$ is a normalisation factor.  In the following this factor will be equal to $\frac{1}{a^2}$ for energy normalisation between the different scales, the studied images and the wavelet transforms at different scales being then in the same unit.

\subsection{Wavelet spectra}
\label{sec:wave_spec}
Following \cite{frick_01} we defined  the energy of the wavelet decomposition map at scale $a$ as:

\begin{equation}
\label{eq_spec}
M(a) = \int_{-\infty}^{\infty}\int_{-\infty}^{\infty} |W(a,\boldsymbol{x})|^2 d\boldsymbol{x}
\end{equation}

where $W(a,\boldsymbol{x})$ is the wavelet decomposition map at the spatial scales $a$.  

The distribution of this energy with the scale $a$ is the wavelet spectrum. This spectrum reveals the dominant scales in the initial image.

\subsection{Wavelet cross-correlation}
\label{sec:wave_corr}
The cross-correlation is a useful method to compare different images. \cite{frick_01} showed that the pixel-to-pixel cross-correlation is dominated by the large scale structures. The information given by the wavelet transformations is used to probe the correlation between different wavelengths, for example, on the scales of the spiral arms, the interarm and central regions. 
The wavelet cross-correlation coefficient at scale $a$ is defined following \cite{frick_01}:

\begin{equation}
\label{eq_corr}
r_w(a) = \frac{\iint W_1(a,\boldsymbol{x}) W_2(a,\boldsymbol{x}) d\boldsymbol{x}}{\left[ M_1(a)M_2(a)\right]^{1/2}}
\end{equation}

where $W_1$ and $W_2$ refer to the wavelet decomposition maps at scale $a$ of two images of same size and geometry and $M_1$ and $M_2$ correspond to the wavelet coefficients at this scale (see Eq.~\ref{eq_spec}),
and the uncertainty on this cross-correlation is defined as:

\begin{equation}
\Delta r_w(a) = \frac{\sqrt{1-r_w^2(a)} }{\sqrt{N-2}}
\end{equation}

with $N=(L/a)^2$, $L$ being the linear size of the image.
 In this analysis we use $r_2=0.75$ as the arbitrary threshold for a significant correlation, following \cite{frick_01} and \cite{fatemeh_07}.

\subsection{Energy conservation}

The wavelet transform is conservative: there is no loss of energy when applying the wavelet transform to an image $f(\boldsymbol{x})$. We indeed have a Plancherel relation for the wavelet transform, similar to the one for the Fourier transform, in the case of an axisymmetric wavelet:

\begin{equation}
\label{eq_energy}
||f||^2=\int f(\boldsymbol{x})^2d\boldsymbol{x} = \dfrac{2\pi}{C_\psi}\int M(a) \dfrac{da}{a}
\end{equation}

where $M(a)$ is the energy of the wavelet transform of $f(\boldsymbol{x})$ at the spatial scales $a$ and $C_\psi=\int{ \dfrac{|\hat{\psi}(\boldsymbol{k})|^2}{k^2} d\boldsymbol{k}}$  a constant,  with $\hat{\psi}$ the wavelet Fourier transform.  This conservation equation is exact for continuous scales from 0 to infinity. Of course, in practice we need to choose a finite numbers of scales on which the wavelet transform is produce. However, Equation~\ref{eq_energy} can be used to quantify the contribution of the structures between  2 spatial scales to the total energy. 

\subsection{Synthesis}

As a consequence of the energy conservation, we have an exact reconstruction of the image $f$ from its wavelet decomposition:

\begin{equation}
\label{eq_synthesis}
 f(\boldsymbol{x})= \dfrac{2\pi}{C_\psi}\int W(a,\boldsymbol{b}) \psi\left( \dfrac{\boldsymbol{x}-\boldsymbol{b}}{a}\right)  d\boldsymbol{b}\dfrac{da}{a}
\end{equation}

with $W(a,\boldsymbol{b})$ being the wavelet transform of $f(\boldsymbol{x})$, with respect to the wavelet $\psi$. As for the energy conservation equation, this reconstruction is mathematically exact but not realizable in practice since it requires an infinity number of scales from 0 to $+\infty$. Image reconstruction (and compression) with wavelets uses the discrete wavelet transform, for which an exact image reconstruction can be achieve with a finite and discreet number of scales. 

\section{Wavelet analysis: effect of the noise}
\label{noise_effect}

\begin{figure}
\begin{center}
\includegraphics[width=8cm]{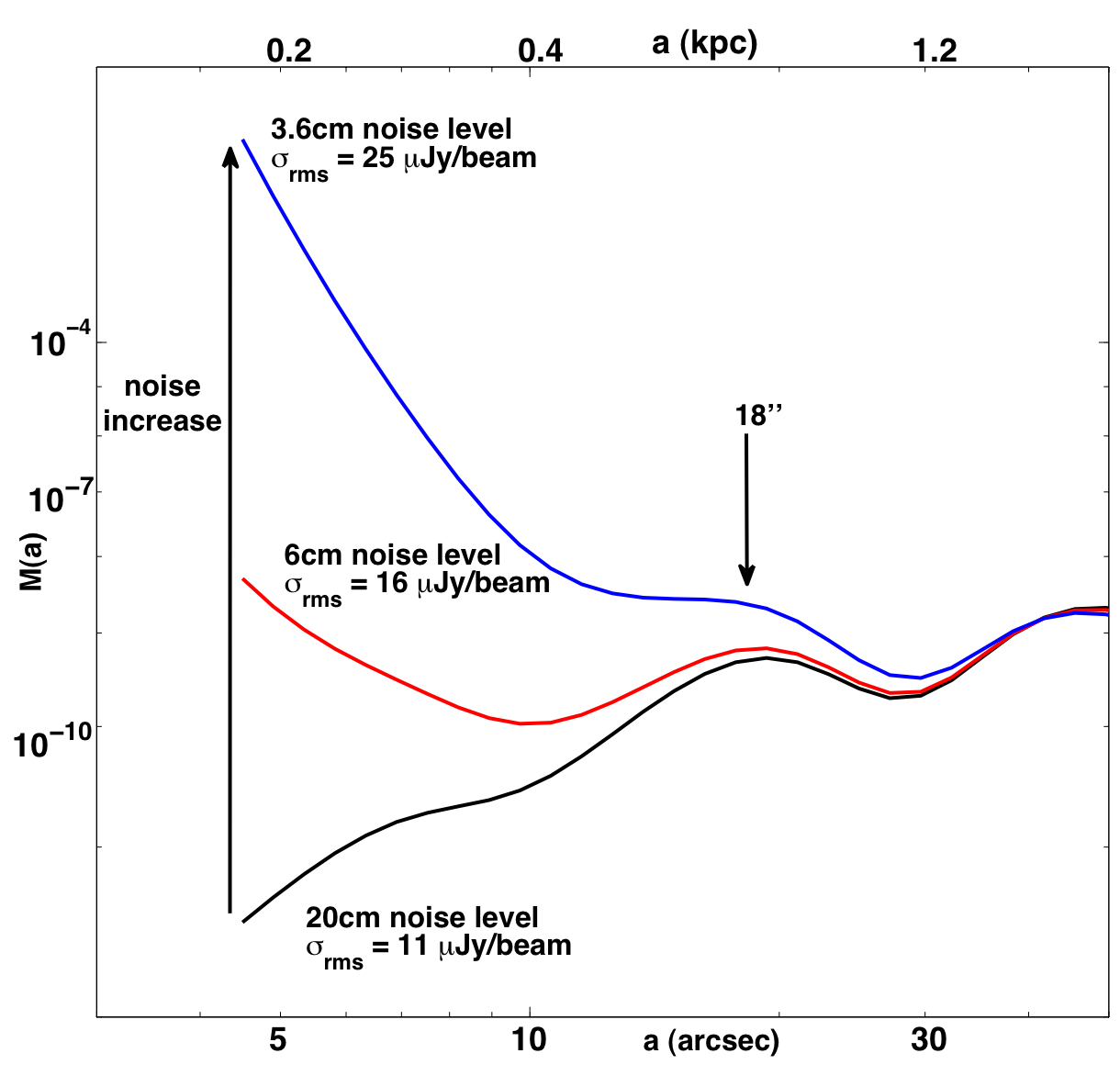} 
\includegraphics[width=8cm]{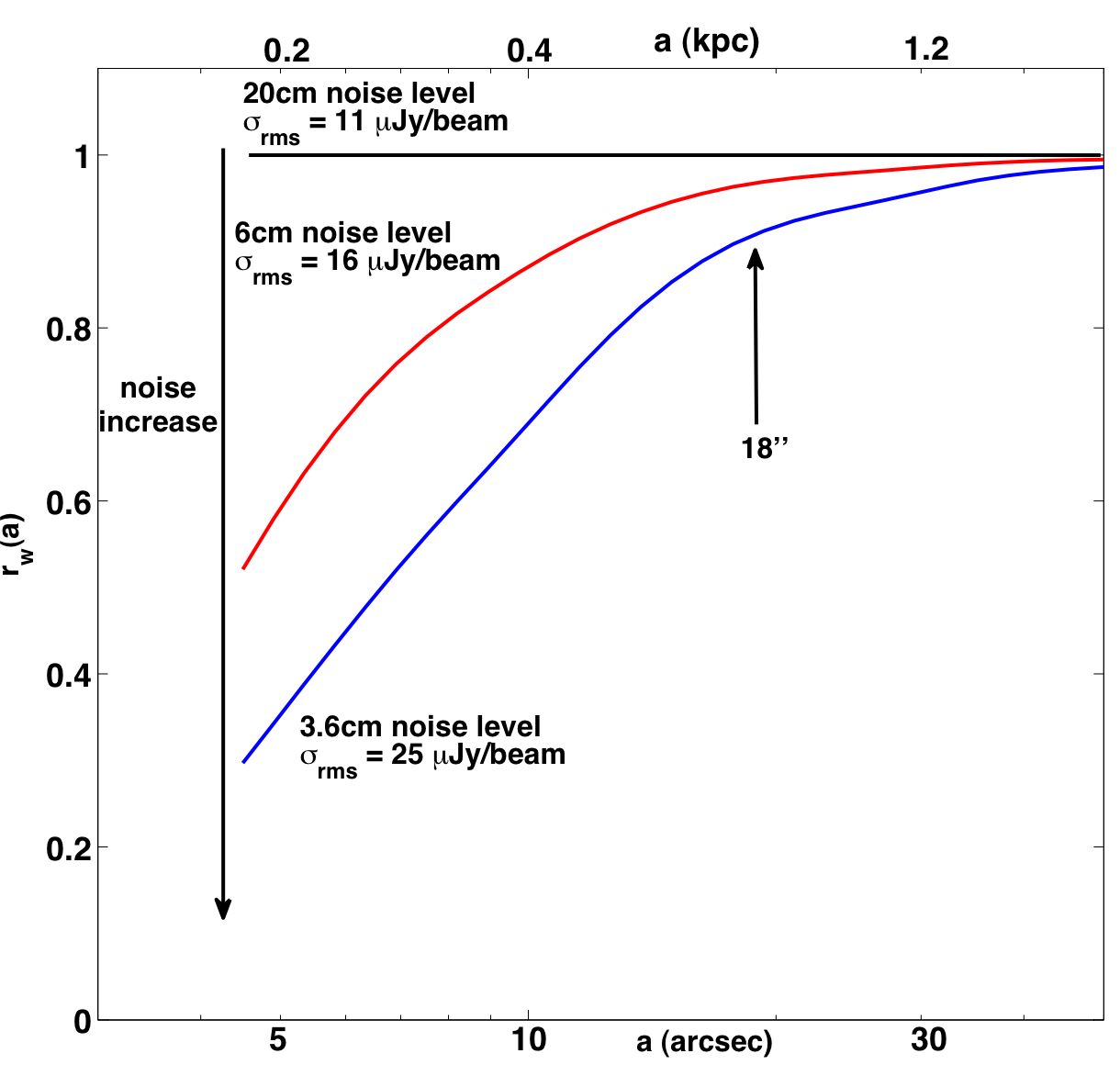} 
\includegraphics[width=8cm]{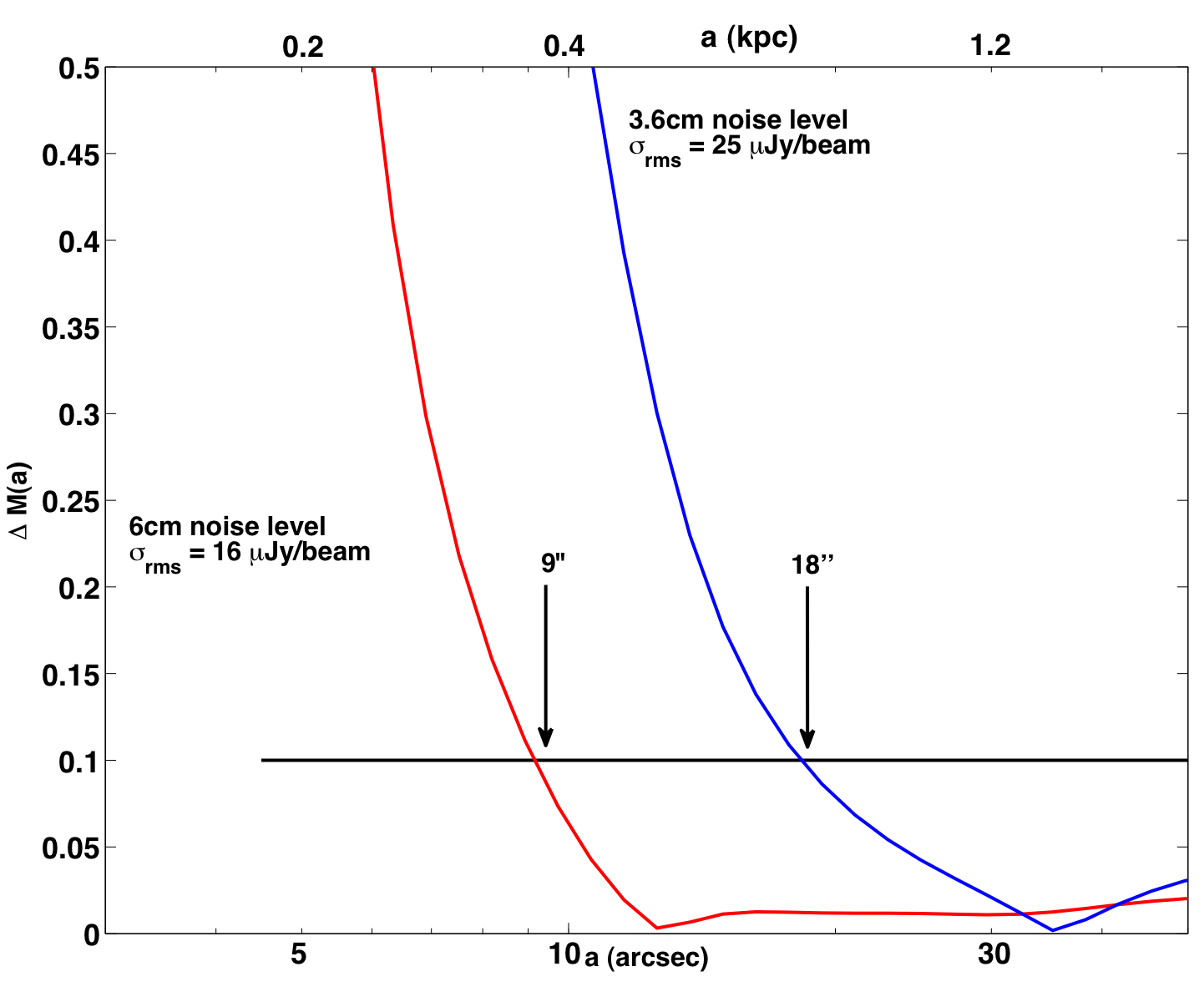} 
\includegraphics[width=8cm]{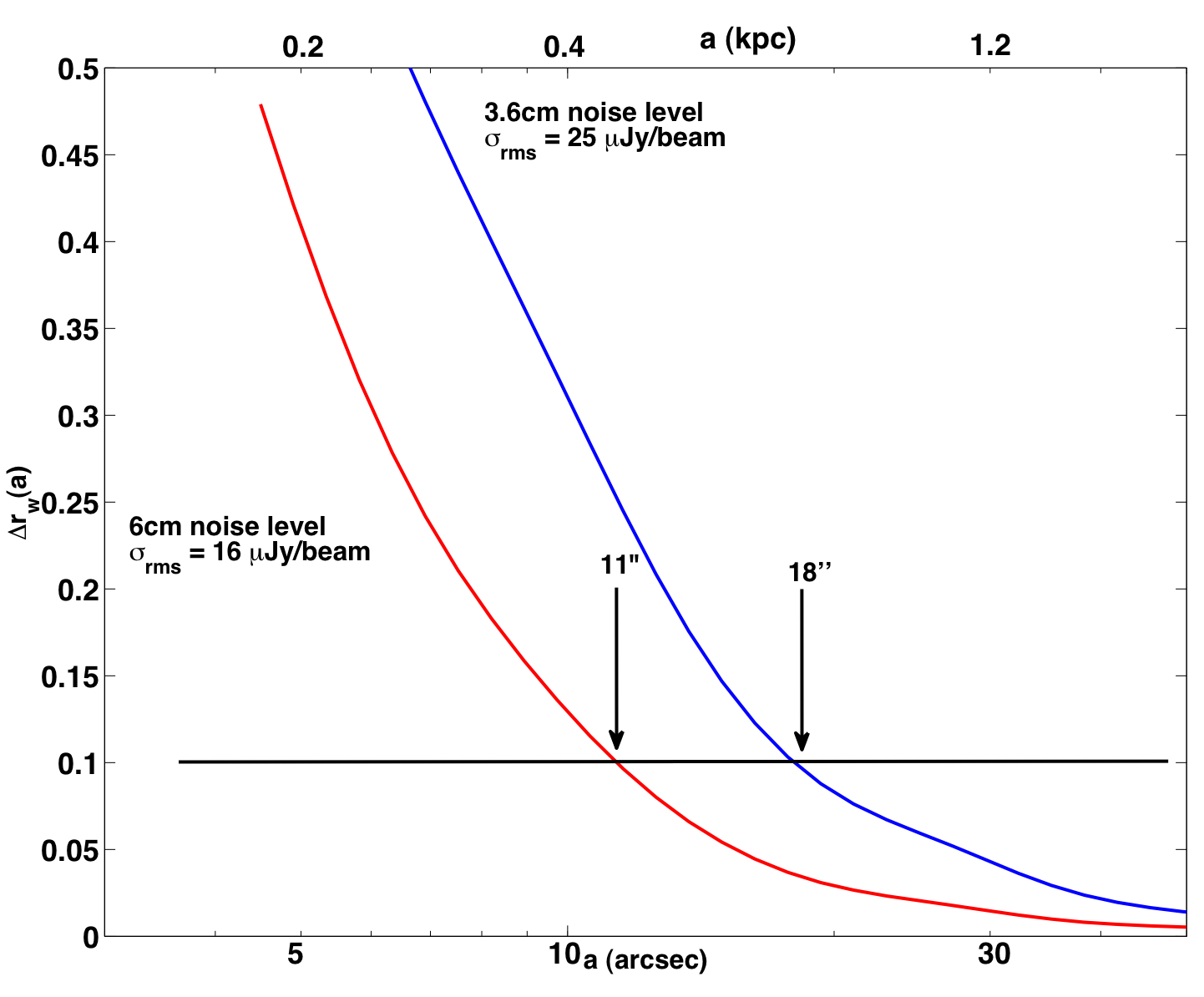} 
\caption{Left: wavelet spectra M(a) of the 20\,cm maps with different noise levels (top) and the relative differences between the spectra of the initial 20\,cm map and each noisy map $\Delta$M(a) (bottom). Right: wavelet cross-correlations r${_w}$ between the initial and noisy 20\,cm data (top) and their relative distance to 1, the perfect correlation, $\Delta$r$_{w}$ (bottom ). The red and blue lines in all four panels correspond to the noise level of the 6\,cm and 3.6\,cm data, respectively. The horizontal line in the two bottom plots corresponds to a relative difference of 10\%.} 
\label{test_noise}
\end{center}
\end{figure}

 We tested the influence of noise by gradually adding Gaussian noise  to our 20   cm map (which has the best signal-to-noise ratio of the radio data)  from 11~$\mu$Jy/beam (noise level of the 20   cm map) to 25~$\mu$Jy/beam (noise level of the 3.6   cm map) in steps of 1~$\mu$Jy/beam.  We then  compute the corresponding wavelet spectra, $M(a)$, for the 20   cm map and each of the noisy maps, and the wavelet cross-correlations  $r_w$ between the 20   cm and noisy maps (see Sect.~\ref{sec:results} and Appendix~\ref{app:wave} for the definitions of the wavelet spectrum and cross-correlation). The result of this test is shown in Figure~\ref{test_noise}, where we plotted only the curves corresponding to the 20   cm , 6   cm and 3.6   cm noise levels, for clarity. We define the relative differences between the initial and noisy maps as $\Delta M(a) = \left|\dfrac{M_{init}(a)-M_{noisy}(a)}{M_{init}(a)} \right| $ and $\Delta r_w = 1-r_{w,noisy}(a)$, where $M_{init}$ ($M_{noisy}$) is the spectrum of the initial 20   cm map (each noisy map)  and  $r_{w,noisy}$ is the wavelet cross-correlation between the initial 20   cm map and each noisy map. The value 1 in the expression of $\Delta r_w$ correspond to the maximum in the correlation. Fig.~\ref{test_noise}, (top left) shows that the wavelet spectra of the noisy maps  differ at small scales,  as the noise affects the spectra by adding energy at those scales. On the other hand, the cross-correlations between the initial and noisy 20   cm maps show that if the noise increases, the cross-correlation decreases rapidly (Fig.~\ref{test_noise}, top right). The relative differences between the spectra and the cross-correlations are larger than 10\% at scales smaller than  18\arcsec\ for the noisiest map, corresponding to the noise level of the 3.6   cm map. At the noise level of the 6   cm map (16$\mu$Jy/beam) the relative differences reach 10\% for scales smaller than 10\arcsec.

\newpage

\begin{deluxetable}{l c c c}
\tabletypesize{\footnotesize}
\tablecolumns{5} 
\tablewidth{\textwidth} 
\tablecaption{Data sets used in this study.}
\startdata
\hline
\hline
$\lambda$ & telescope & resolution &rms  \\
           &            & [\arcsec$\times$\arcsec]   &     \\
\hline
20\,cm&VLA (ABCD)& 1.4$\times$1.3&11\,$\mu$Jy/beam  \\

6\,cm&VLA (BCD) + Effelsberg& 2.0$\times$2.0& 16\,$\mu$Jy/beam \\
3.6\,cm&VLA (CD) + Effelsberg& 2.4$\times$2.4&25\,$\mu$Jy/beam\\
24\,$\mu$m\,\tablenotemark{a}&SST/MIPS HiRes& 2.4$\times$2.3& 0.0067\,MJy/sr \\
8\,$\mu$m PAH\,\tablenotemark{b}&SST/IRAC HiRes& 1.3$\times$1.1&  0.0069\,MJy/sr\\
H$\alpha$\,\tablenotemark{c}& 2.1m KPNO& 1.9$\times$1.9&$1.8\times 10^{-17}$\,{\small erg\,s$^{-1}$\,cm$^{-2}$\,arcsec$^{-2}$} \\
\enddata
\label{data}
\tablenotetext{a}{ data from the SINGS DR5}
\tablenotetext{b}{ data from \cite{regan_pah06}}
\tablenotetext{c}{ data from \cite{sings_2005}}
\end{deluxetable}

\begin{deluxetable}{l c c c c c c }
\tablecaption{Summary of VLA observations.}
\tabletypesize{\footnotesize}
\tablewidth{20cm}
\rotate
\startdata
\hline
\hline
$\lambda$ &  configuration & central frequencies & Bandwidth&project ID&  date & on-source integration time \\
          &                &[GHz]                &  [MHz]      &       & & [h] \\
\hline
3.6\,cm & C &8.4351;8.4851&50&AS780  & '04 Apr 05-06 & 20.3\\
      &  D &8.4351;8.4851&50&AB999  & '01 Oct 28-29& 17.6\\
\hline
   6\,cm & B&4.8351;4.8851&25&AS780  & '03 Dec -'04 Jan& 23\\
       &  C  &4.8351;4.8851&50&AB999 & '01 Aug 18-26 & 14.4\\
       &   D &4.8351;4.8851&50&AB591 & '91 Mar 24 -Apr 01& 8.7\\
\hline
  20\,cm& A&1.3851;1.4649&25&AS723,AS812& '02 Apr 05 - '04 Oct-Nov & 30.8\\
       &  B  &1.3851;1.4649&25&AS812& '05 June 1 \& 5-6 & 23.9\\
       &   C &1.4649;1.5149&50&AN57  & '92 Mar 30 & 13.4\\
       &   D &1.4649;1.6649&50&AB505 & '88 Aug 27& 12\\

\enddata
\label{vladata_obs}
\end{deluxetable}

\begin{deluxetable}{l | c c c}
\tablecaption{Radio properties of M51.}
\startdata
\hline
\hline
$\lambda$ &20\,cm & 6\,cm &3.6\,cm  \\
\hline
$S_\nu$ (Jy)&$1.4\pm0.1$&$0.6\pm0.2$&$0.3\pm0.1$\\
$h$ (kpc)&4.5$\pm$0.5&5.6$\pm$0.4&3.7$\pm$0.2\\
$I_0$ ($\mu$Jy/beam)&55$\pm$9&30$\pm$4&37$\pm$5
\label{radio_prop}
\enddata
\tablecomments{Values were measured using the task IRING in AIPS. The total flux $S_\nu$ is measured out to a radius of 400\arcsec. The scale lengths $h$ and origin point $I_0$ come from fits to the radial profiles presented in Fig.~\ref{fig:radial_profile} with exponentials of the form $I_0\,\times\,exp\left( -\frac{r}{h}\right)$.}
\end{deluxetable}

\begin{deluxetable}{r|l|p{2.5cm}|p{2.5cm}|p{2.5cm}|l}
\tabletypesize{\footnotesize}
\tablewidth{20cm}
\tablecaption{Interesting spatial scales from the wavelet analysis of M51a.}
\tablehead{\colhead{$a$ [kpc]} &\colhead{MS} & \colhead{mS}  &\colhead{MX}  &\colhead{mX}  & \colhead{Comments}}
\rotate
\startdata
0.6$\pm$0.05 & 24; 8; Ha & - &  6/Ha & - & \HII\,complexes\\ \hline
0.8$\pm$0.08  & 20; 6; 3.6 & - & - & - &  small structures in M51a disk:  \\ \cline{1-5}
0.9$\pm$0.1 & - & 8 & - & 20/24;6/Ha & \HII\ regions, PDR... \\ \hline
1.2$\pm$0.2 & - & 20; 6; 3.6 & - &-  & filamentary structures \\ \hline
2.0$\pm$0.2 & 20; 6; 3.6; 8 & - & radio/radio; radio/24;20-6/Ha & - & \parbox[t][2.cm]{6.5cm}{width of the spiral arms, wider at 20cm (2.2kpc) than at the other wavelengths (1.8kpc), due to CRes diffusion} \\ \hline
3.6$\pm$0.4 & - & 20; 6; 3.6; 24; 8 & - &radio/radio; radio/24&  best arm/interarm contrast \\ \hline
4.5$\pm$0.4 & 24; 8 & - & - & - &  M51a inner disk, larger in radio than \\ \cline{1-5}
5.7$\pm$0.6& 20; 6; 3.6 & - & radio/Ha; IR/Ha; 8/24 & - &  in IR and optical images \\ \hline
7.3$\pm$0.7&- & - &20-6/3.6; 3.6/24 & 8/24 &  \\ \hline
8.4$\pm$0.8&- & 20; 6; Ha; 24; 8 &- & radio/Ha; IR/Ha; 3.6/24 & M51a + interaction region \\ \hline
10.3$\pm$1.1&- & - &- & 20-6/3.6 3.6/24 &  \\ \hline
22$\pm$2.1&24; 8; Ha & - &- & - & M51a galactic disk, larger in radio that   \\ \cline{1-5}
$>$25&20; 6; 3.6 & - &- & - &  in IR and optical emission
\label{wave_minmax}
\enddata
\tablecomments{The first column lists the spatial scales $a$ in kpc, corresponding to local extrema in the wavelet spectra and cross-correlations. The errors on $a$  are the maximum between the steps $(\sim\,\frac{a}{10})$ and the spatial shift of the local extrema  between the different wavelengths.   
The following four columns indicate these extrema: 'MS' and 'mS' stand respectively for maximum and minimum in the wavelet spectra, and 'MX' and 'mX' for maximum and minimum in the cross-correlations. The numbers indicate the different wavelengths: '20' for 20\,cm, '6' for 6\,cm, '3' for 3.6\,cm, '24' for 24\,$\mu$m, '8' for 8\,$\mu$m and 'Ha' for H$\alpha$. The cross-correlation between $\lambda_1$ and $\lambda_2$  is noted as : '$\lambda_1$/$\lambda_2$'.}
\end{deluxetable}

\begin{deluxetable}{c c c c c c}
\tablecaption{Results of the fit: $F_{24\mu m} = b (S_{20   cm})^c$}
\tablehead{\colhead{fit parameters} & \colhead{spiral arms} &\colhead{interarm} & \colhead{outer}  &\colhead{inner}  &\colhead{total}  }

\startdata
$\langle c\rangle$         &1.01           &0.51     &0.48   &1.5    &1.00\\
$\sigma_{c}$                  &0.01           &0.01     &0.01   &0.1    &0.01\\
$\langle \log{b}\rangle$&0.9           &-0.50   &-1.19 &2.3    &0.81\\
$\sigma_{\log{b}}$         &0.01           &0.01     &0.02   &0.3   &0.03\\
\hline
$\langle q_{24}\rangle$&1.01          &0.91  &0.63 &1.0    &0.86 \\
$\sigma_{q_{24}}$         &0.2           &0.16     &0.17   &0.1   &0.3
\label{fit_values}
\enddata
\tablecomments{Means and dispersions of the fit parameters $b$ and $c$ for the different regions, assuming $F_{24\,\mu m} = b (S_{20\,cm})^c$ (fluxes derived within 6\arcsec\ apertures). The means and dispersions of the  $q_{24}$ parameter from \cite{murphy_06} are also listed for comparison.}
\end{deluxetable}

\label{lastpage}

\end{document}